\documentclass[12pt]{article}
\usepackage{amsfonts,amsthm,amsmath,amssymb,upgreek,bm}
\usepackage{hyperref}
\usepackage[paper=letterpaper,margin=.85in]{geometry}
\usepackage{graphicx}
\usepackage{units}
\usepackage[dvipsnames]{xcolor}
\usepackage{float}
\usepackage[export]{adjustbox}
\usepackage[dvipsnames]{xcolor}
\usepackage[shortlabels]{enumitem}
\usepackage[mathscr]{euscript}
\usepackage[normalem]{ulem}
\usepackage{bm}
\usepackage{tikz}
\usepackage{tikzsymbols}
\usepackage{pgfplots}
\usepackage{ytableau}
\usepackage{slashed}
\usepackage{mathtools}
\usepackage{soul}
\sethlcolor{yellow}
\soulregister\ref7
\soulregister\eqref7
\usepackage{bbold}
\usetikzlibrary{decorations.pathmorphing}
\usetikzlibrary{arrows.meta}
\usetikzlibrary{calc}
\usepgfplotslibrary{fillbetween}
\usepackage[nameinlink]{cleveref}
\usepackage{mathrsfs}
\usepackage[bbgreekl]{mathbbol}

\newcommand{\dd}{\mathrm d}

\newcommand{\be}{\begin{equation}}
\newcommand{\ee}{\end{equation}}
\newcommand{\bea}{\begin{eqnarray}}
\newcommand{\eea}{\end{eqnarray}}

\numberwithin{equation}{section}

\begin{document}
\thispagestyle{empty}

\vspace*{2.5cm}
\begin{center}

{\bf {\LARGE Small black hole in de Sitter as an observer}}

\begin{center}

\vspace{1cm}

{Guanda Lin${}^1$ and Haifeng Tang${}^2$}\\
 \bigskip \rm

\bigskip

{${}^1$  Leinweber Institute for Theoretical Physics and Department of Physics, University of California, Berkeley, California 94720, U.S.A.}
\,\vspace{0.5cm}\\
{${}^2$ Leinweber Institute for Theoretical Physics, Stanford University, Stanford, CA 94305, USA}

\rm
  \end{center}

\vspace{2.5cm}
{\bf Abstract}
\end{center}
\begin{quotation}
\noindent
We study the Euclidean path integral around magnetically charged lukewarm black holes in four-dimensional de Sitter space. We numerically determine the full one-loop spectrum of coupled graviton and photon fluctuations and find that the phase of the partition function agrees with that of pure gravity in empty de Sitter plus a worldline observer.
In the small-mass limit, certain localized eigenmodes can be effectively viewed as describing the  observer's motion and have the expected eigenvalues, allowing us to reconstruct the worldline action.
We further identify these modes with translational zero modes of an extremal charged black hole in asymptotically flat spacetime, lifted by de Sitter curvature.

\end{quotation}

\setcounter{page}{0}
\setcounter{tocdepth}{2}
\setcounter{footnote}{0}

\newpage

\setcounter{page}{2}
\tableofcontents

\newpage

\section{Introduction}
\label{sec:introduction}
In $D$ spacetime dimensions, the round sphere $S^D$ is a solution of Euclidean gravity with positive cosmological constant, and its classical action is minus the de Sitter entropy~\cite{Gibbons:1976ue,Gibbons:1977mu}, suggesting that the sphere path integral counts states in de Sitter space.

A difficulty with this interpretation is Polchinski's result for the graviton one-loop partition function $Z_{S^D}$: its phase is $\mathrm{i}^{D+2}$~\cite{Polchinski:1988ua}, which is not generically real and positive as expected for a count of states.
Two facts account for this phase.
First, most modes of the conformal factor have negative quadratic action, so their integrals require a Wick rotation~\cite{Gibbons:1978ac}.
Second, $D+2$ conformal modes do not need this rotation.
They belong to the $\ell_0=0$ and $\ell_0=1$ sectors, where $\ell_0$ denotes scalar angular momentum on $S^D$.
The $\ell_0=0$ mode changes the overall size of the sphere and is a stable physical fluctuation.
The $\ell_0=1$ modes are pure-gauge zero modes before gauge fixing, corresponding to the $D+1$ conformal Killing vectors that are not Killing vectors.
The gauge-fixing term makes them positive; alternatively, these gauge directions can be omitted from the integral~\cite{Law:2020cpj,Law:2025yec}.
In either treatment, the formal phase is $(-\mathrm{i})^{\infty-1-(D+1)}$, which becomes $\mathrm{i}^{D+2}$ after absorbing the formal factor $(-\mathrm{i})^\infty$ into the definition of an ultralocal measure.

Maldacena proposed to resolve this puzzle by including an observer~\cite{Maldacena:2024spf}.\footnote{Observers \cite{Chandrasekaran:2022cip,Witten:2023xze,Witten:2026whn} play an important role in recent understanding of de Sitter holography. See also observer path integrals and Hilbert spaces~\cite{Abdalla:2025gzn,Harlow:2025pvj}, and de Sitter out-of-time-order correlators~\cite{Kolchmeyer:2024fly,Milekhin:2026tbi,Cui:2026bcd,Chen:2026boh,Harlow:2026pwe}.}
Of the $D+1$ conformal Killing directions, all but two then become physical because they move the observer.
Equivalently, the observer has $D-1$ unstable position modes that must be Wick-rotated.
Either description gives the phase $\mathrm{i}^{D+2}(-\mathrm{i})^{D-1}=\mathrm{i}^3=-\mathrm{i}$.\footnote{Further work studied the phase for product manifolds~\cite{Ivo:2025yek,Shi:2025amq, Ivo:2025xek}, the Coleman--de Luccia instanton~\cite{Ivo:2025fwe}, de Sitter axion wormholes~\cite{Ivo:2026ijv}, and higher spin gravity~\cite{Giombi:2026sqa}. }

Reference~\cite{Chen:2025jqm} studied this proposal with a charged black hole as the observer, within the framework of dimensionally reduced two-dimensional dilaton gravity.
The black hole is a solution of the gravitational theory itself, so its backreaction is included without adding a separate worldline to the action.
In four-dimensional Einstein--Maxwell theory, the lukewarm Reissner--Nordstr\"om--de Sitter solutions have equal black hole and cosmological horizon temperatures and admit smooth Euclidean continuations.
The reduced calculation found the phase $-\mathrm{i}$ along this family.
By imposing the Hamiltonian constraint through an inverse Laplace contour, before performing the conjugate energy integral, they obtain  a positive density of states for the stable lukewarm saddle in the reduced theory.\footnote{See also perspectives from Lorentzian gravitational path integrals~\cite{Marolf:2022ybi,Kolanowski:2026gii,Marolf:1996gb,Casali:2021ewu}.}

Two limitations of that calculation motivate the present work.
Most importantly, the cancellation of the $D$-dependent phase involves the $D-1$ conformal Killing vectors
that displace the observer.
They belong to the
$\ell=1$ sector under $SO(D-1)$, where $\ell$ denotes angular momentum on the transverse $S^{D-2}$.
They are therefore absent from a reduction restricted to the spherically symmetric $\ell=0$ sector in the previous work \cite{Chen:2025jqm}.
In addition, the reduced calculation fixed the Maxwell field to its background value, whereas the full quadratic operator couples metric and Maxwell fluctuations.

In this work, we study the full spectrum of these coupled fluctuations on the
$D = 4$
lukewarm black hole background with general mass.
Our first result is that the one-loop phase remains $\mathrm{i}^3$ along the lukewarm family, in agreement with Maldacena's worldline computation.
We decompose the fluctuations under $U(1)\times SO(3)$, and
label the decomposed sectors
with Euclidean-time Fourier number $n$ and angular momentum $\ell$, respectively.
The nonzero regularized counts occur only at $(n,\ell)=(0,0)$ and $(\pm 1,0)$, giving a total of $-3$ and hence the phase $(-\mathrm{i})^{-3}=\mathrm{i}^3$.
In particular, the $(0,1)$ sector contributes no net phase.
This is the sector containing the three sphere conformal Killing directions that displace the observer.
The full calculation thus tests the observer's cancellation of the $D$-dependent phase, including backreaction, in $D=4$.

Our second result identifies the observer's motion directly in the one-loop fluctuation spectrum.
In the limit of small black hole mass, we  find the eigenfunctions are either extended to the cosmological horizon or localized near the black hole horizon.
The extended family approaches the eigenfunctions of gravitons and photons on empty de Sitter,
while the localized family appears only in the $\ell=1$ sector and is supported at radii of order the black hole horizon radius.
Their support approaches $r=0$, the center of the de Sitter static patch, while their eigenvalues remain finite in de Sitter units.
Their dependence on $n$ agrees with the spectrum of a massive worldline particle in empty de Sitter.
We explain these properties by relating the localized eigenfunctions to the translation zero mode of an extremal charged black hole in asymptotically flat spacetime.
De Sitter curvature
lifts this zero mode.
We calculate the resulting eigenvalues and the norm of a mode with specified displacement, recovering the worldline action with the physical black hole mass as its coefficient.
The small black hole thus reproduces both the phase and the dynamics expected of an observer.

Section~\ref{sec:one-loop} first reviews the lukewarm black hole solution, sets up the one-loop spectrum calculation, and then presents the numerical spectrum and discusses the phase and the small-mass limit.
Section~\ref{sec:translation-worldline} explains the localized modes and reconstructs the worldline action from them.
Technical derivations are collected in the appendices.

\section{One-loop fluctuations of the magnetic black hole}
\label{sec:one-loop}

\subsection{Background and one-loop setup}
\label{sec:background-one-loop}
We first review the lukewarm black hole as a solution to four-dimensional Einstein--Maxwell theory and then formulate the coupled one-loop spectrum problem on this classical background.
We keep Newton's constant $G_N$ explicit.
Let $l_\text{dS}$ be the de Sitter radius and $M$ the physical black hole mass, so $G_NM$ has units of length.
We use the dimensionless mass
\begin{equation}
 \mathsf M\equiv\frac{G_NM}{l_\text{dS}}.
 \label{eq:dimensionless-black hole-mass}
\end{equation}
The phase calculation concerns the lukewarm family $0< \mathsf M<1/4$, while later the worldline description applies specifically for $\mathsf M\to0$.

We work in four-dimensional Euclidean Einstein--Maxwell theory with cosmological constant $\Lambda=3/l_\text{dS}^2$.
For a metric $g_{\mu\nu}$ and a Maxwell potential $A_\mu$ with field strength $F_{\mu\nu}$, the action is
\begin{equation}
 I[g,A]=\frac{1}{16\pi G_N}\int\dd^4x\,\sqrt g
 \left(-R+2\Lambda+F_{\mu\nu}F^{\mu\nu}\right).
 \label{eq:einstein-maxwell-action}
\end{equation}
We use the normalization in which the Maxwell term has the same overall factor as the Einstein term.

An overbar denotes a background field, and $Q$ is the magnetic charge, with units of length.
Writing $\dd\Omega_2^2=\dd\theta^2+\sin^2\theta\,\dd\varphi^2$ for the unit two-sphere metric, the magnetic Reissner--Nordstr\"om--de Sitter (RN--dS) solution in Euclidean signature is
\begin{equation}
 \begin{aligned}
 \dd\bar s^2&=f(r)\dd\tau^2+\frac{\dd r^2}{f(r)}+r^2\dd\Omega_2^2,\ \ \ \ \ \ f(r)=1-\frac{2G_NM}{r}+\frac{Q^2}{r^2}-\frac{r^2}{l_\text{dS}^2},\\
 \bar F&=Q\sin\theta\,\dd\theta\wedge\dd\varphi.
 \end{aligned}
 \label{eq:lukewarm-background}
\end{equation}
The magnetic solution has a real Euclidean field strength~\cite{Mann:1995vb}, which makes it convenient for the fluctuation calculation.
The Euclidean region lies between the outer black hole horizon $r_h$ and the cosmological horizon $r_c$, where $f$ is positive and has a simple zero at each endpoint.
Let $\beta$ be the period of $\tau$, and let a prime denote differentiation with respect to $r$.
The thermal circle closes smoothly at both horizons when their temperatures agree, so that
\begin{equation}
 \beta=\frac{4\pi}{f'(r_h)}
      =\frac{4\pi}{|f'(r_c)|},
 \qquad \tau\sim\tau+\beta,
 \label{eq:common-euclidean-period}
\end{equation}
On the lukewarm family this condition gives~\cite{Romans:1991nq,Booth:1998gf,Morvan:2022aon}
\begin{equation}
 |Q|=G_NM=l_\text{dS}\mathsf M.
 \label{eq:lukewarm-charge}
\end{equation}
The metric function, horizon radii, and common period are then
\begin{equation}
 \begin{aligned}
 f(r)&=\left(1-\frac{l_\text{dS}\mathsf M}{r}\right)^2-\frac{r^2}{l_\text{dS}^2},\\
 r_h&=\frac{l_\text{dS}}{2}\left(1-\sqrt{1-4\mathsf M}\right),
 &r_c&=\frac{l_\text{dS}}{2}\left(1+\sqrt{1-4\mathsf M}\right),\\
 \beta&=\frac{2\pi l_\text{dS}}{\sqrt{1-4\mathsf M}}.
 \end{aligned}
 \label{eq:lukewarm-parameters}
\end{equation}
The endpoints of the lukewarm line $0<\mathsf M<\frac14$ are limiting geometries: the black hole horizon shrinks as $\mathsf M\to0$, while the two horizons coincide in the Nariai limit $\mathsf M\to1/4$; see Ref.~\cite{Castro:2022cuo} for the near-extremal limits of charged de Sitter black holes.
For $0<\mathsf M<1/4$, the Euclidean manifold is topologically a smooth $S^2\times S^2$~\cite{Mann:1995vb}.

We now expand about this background to set up the one-loop calculation.
We denote the metric and gauge field perturbations by $h_{\mu\nu}$ and $a_\mu$.
At each background we hold the magnetic flux fixed, but allow the Maxwell field to fluctuate:
\begin{equation}
 g_{\mu\nu}=\bar g_{\mu\nu}+h_{\mu\nu},
 \qquad A_\mu=\bar A_\mu+a_\mu,
 \qquad F=\bar F+\dd a.
 \label{eq:coupled-fluctuations}
\end{equation}
Although $\bar A$ requires gauge patches, $a$ is a globally defined one-form in the fixed magnetic-flux sector.
The nonzero background field strength $\bar F$ couples the metric and Maxwell fluctuations in the quadratic action, so freezing $a_\mu$ would not give the full spectrum.

The metric trace is $h=\bar g^{\mu\nu}h_{\mu\nu}$, the trace-reversed perturbation is $\widetilde h_{\mu\nu}=h_{\mu\nu}-\frac12\bar g_{\mu\nu}h$, and the field-strength perturbation is $f_{\mu\nu}=\nabla_\mu a_\nu-\nabla_\nu a_\mu$.
Overbars also denote background curvature tensors.
The Einstein--Hilbert and Maxwell contributions to the quadratic action, denoted by $I_{\rm EH}^{(2)}$ and $I_{\rm M}^{(2)}$, are, up to integrated total derivatives,
\begin{equation}
 \begin{aligned}
 I_{\rm EH}^{(2)}
 =\frac{1}{16\pi G_N}\int\dd^4x\,\sqrt{\bar g}\,
 \bigg\{&
 \frac14\widetilde h^{\mu\nu}
 \Big[-\nabla^2h_{\mu\nu}
       -2\bar R_{\rho(\mu}h_{\nu)}{}^\rho
       -2\bar R_{\mu\rho\nu\sigma}h^{\rho\sigma}
       +(\bar R-2\Lambda)h_{\mu\nu}\Big]\\
 &-\frac12
  \big(\nabla^\rho\widetilde h_{\rho\mu}\big)
  \big(\nabla^\sigma\widetilde h_\sigma{}^\mu\big)
 \bigg\},
 \end{aligned}
 \label{eq:quadratic-einstein-action}
\end{equation}
and
\begin{equation}
 \begin{aligned}
 I_{\rm M}^{(2)}
 =\frac{1}{16\pi G_N}\int\dd^4x\,\sqrt{\bar g}\,
 \bigg[&
 \bar F_{\alpha\beta}\bar F^{\alpha\beta}
     \left(-\frac14h_{\mu\nu}h^{\mu\nu}+\frac18h^2\right)
 -h\,\bar F_{\mu\rho}\bar F_\nu{}^\rho h^{\mu\nu}\\
 &+\bar F_{\mu\nu}\bar F_{\rho\sigma}h^{\mu\rho}h^{\nu\sigma}
 +2\bar F_{\mu\rho}\bar F_\nu{}^\rho
          h^{\mu\sigma}h^\nu{}_\sigma+f_{\mu\nu}f^{\mu\nu}
 \\
 &+h\,\bar F^{\mu\nu}f_{\mu\nu}
 -4\bar F_{\mu\nu}f_{\rho\sigma}\bar g^{\mu\rho}h^{\nu\sigma}\bigg].
 \end{aligned}
 \label{eq:quadratic-maxwell-action}
\end{equation}
The terms in the third line of \eqref{eq:quadratic-maxwell-action} couple the metric and Maxwell fluctuations.\footnote{A related expansion of the coupled Einstein--Maxwell action and its gauge-fixing terms is given in Ref.~\cite{Karan:2021teq}, with different fluctuation normalizations and vanishing cosmological constant.}

We use de Donder gauge for the metric and Lorenz gauge for the Maxwell field.
The quadratic gauge-fixing action is\footnote{Variations of the gauge-fixing parameters leave the regularized phase count unchanged in the sectors tested; see Appendix~\ref{app:active-radial-components}.}
\begin{equation}
 I_{\rm gf}^{(2)}
 =\frac{1}{16\pi G_N}\int\dd^4x\,\sqrt{\bar g}\,
 \left[
 \frac12\big(\nabla^\rho\widetilde h_{\rho\mu}\big)
             \big(\nabla^\sigma\widetilde h_\sigma{}^\mu\big)
 +2\big(\nabla^\mu a_\mu\big)^2
 \right].
 \label{eq:coupled-gauge-fixing}
\end{equation}
We denote the full gauge-fixed quadratic action by $I^{(2)}$:
\begin{equation}
 I^{(2)}=I_{\rm EH}^{(2)}+I_{\rm M}^{(2)}+I_{\rm gf}^{(2)}.
 \label{eq:full-quadratic-action}
\end{equation}
The eigenvalues below refer to this gauge-fixed quadratic form.

Write $u=(h_{\mu\nu},a_\mu)$ for a coupled perturbation.
For two perturbations $u_1$ and $u_2$, we use the positive ultralocal inner product
\begin{equation}
 \langle u_1,u_2\rangle_{\rm ultra}
 =\frac{1}{8\pi}\int\dd^4x\,\sqrt{\bar g}
 \left(\frac14h_1^{\mu\nu}h_{2\mu\nu}
                    +2a_1^\mu a_{2\mu}\right).
 \label{eq:coupled-ultralocal-inner-product}
\end{equation}
The squared norm is $\|u\|_{\rm ultra}^2=\langle u,u\rangle_{\rm ultra}$.
Let $\mathcal O_{\rm Lichnerowicz}$ denote the dimensionful coupled fluctuation operator defined by $I^{(2)}$ and this inner product.
We denote the operator in de Sitter units by $\mathcal O$ and its dimensionless eigenvalues by $\lambda$:
\begin{equation}
 \mathcal O\equiv l_\text{dS}^2\mathcal O_{\rm Lichnerowicz},
 \qquad \mathcal O u=\lambda u,
 \qquad
 I^{(2)}[u]=\frac{1}{2G_Nl_\text{dS}^2}
             \langle u,\mathcal O u\rangle_{\rm ultra}.
 \label{eq:dimensionless-fluctuation-operator}
\end{equation}
An eigenmode therefore contributes $\lambda\|u\|_{\rm ultra}^2/(2G_Nl_\text{dS}^2)$ to the action.

\paragraph{Symmetry decomposition.}
The background has $U(1)\times SO(3)$ isometry, corresponding to translations in Euclidean time and rotations of the two-sphere.
We decompose the fluctuations into harmonics of these symmetries, with nonnegative time Fourier number $n$ and angular momentum $\ell$, and then resolve the discrete symmetries.
For the angular dependence, we use the standard scalar- and vector-harmonic construction of Ref.~\cite{Kodama:2003kk}.
Lowercase Latin indices $a,b$ take the values $\tau,r$, and Latin indices $i,j$ denote angular directions.
On the unit two-sphere, let $\Omega_{ij}$ be the metric, $\widehat{\nabla}_i$ its covariant derivative, and $\varepsilon_{ij}$ its volume form.
Choose real orthonormal scalar harmonics $\mathbb{Y}_{\ell m}$, with $\ell\geq0$ and $m$ labeling the $2\ell+1$ harmonics.
Define the transverse vector harmonics $\mathbb{V}_i^{\ell m}$ by
\begin{equation}
 -\widehat{\nabla}^i\widehat{\nabla}_i \mathbb{Y}_{\ell m}=\ell(\ell+1)\mathbb{Y}_{\ell m},
 \qquad
 \mathbb{V}_i^{\ell m}=\varepsilon_i{}^j \widehat{\nabla}_j \mathbb{Y}_{\ell m},
 \qquad \widehat{\nabla}^i \mathbb{V}_i^{\ell m}=0.
 \label{eq:angular-harmonics}
\end{equation}
The vector harmonics exist for $\ell\geq1$.
There are no independent transverse traceless tensor harmonics on $S^2$; the angular tensor components are obtained from scalar and vector harmonics.

Within a fixed $(\ell,m)$ sector of $SO(3)$, denote the scalar-type metric amplitudes by $h_{ab}^{(\mathsf S)}$, $h_{a}^{(\mathsf S)}$, $h_{1}^{(\mathsf S)}$, and $h_{2}^{(\mathsf S)}$, and the vector-type metric amplitudes by $h_{a}^{(\mathsf V)}$ and $h^{(\mathsf V)}$.
The Maxwell amplitudes are $a_{a}^{(\mathsf S)}$ and $a^{(\mathsf S)}$ in the scalar-type expansion, and $a^{(\mathsf V)}$ in the vector-type expansion.
All these amplitudes are functions of $(\tau,r)$.
The decomposition is
\begin{equation}
 \begin{aligned}
 h_{ab}&=h_{ab}^{(\mathsf S)}\mathbb{Y}_{\ell m},\\
 h_{ai}&=h_{a}^{(\mathsf S)}\widehat{\nabla}_i \mathbb{Y}_{\ell m}+h_{a}^{(\mathsf V)}\mathbb{V}_i^{\ell m},\\
 h_{ij}&=h_{1}^{(\mathsf S)}
       \left(\widehat{\nabla}_i\widehat{\nabla}_j+\frac{\ell(\ell+1)}2\Omega_{ij}\right)\mathbb{Y}_{\ell m}
       +\frac12h_{2}^{(\mathsf S)}\Omega_{ij}\mathbb{Y}_{\ell m}
       +h^{(\mathsf V)} \widehat{\nabla}_{(i}\mathbb{V}_{j)}^{\ell m},\\
 a_a&=a_{a}^{(\mathsf S)}\mathbb{Y}_{\ell m},\\
 a_i&=a^{(\mathsf S)} \widehat{\nabla}_i \mathbb{Y}_{\ell m}+a^{(\mathsf V)} \mathbb{V}_i^{\ell m}.
 \end{aligned}
 \label{eq:full-angular-decomposition}
\end{equation}

The background magnetic field is invariant under $SO(3)$ but changes sign under angular parity.
Only the metric--photon mixing terms, in the third line of \eqref{eq:quadratic-maxwell-action}, contain one factor of $\bar F_{ij}$, proportional to $\varepsilon_{ij}$; the metric--metric and photon--photon terms contain an even number of background field strengths.
Consequently, scalar-type metric perturbations couple to vector-type photon perturbations, and vector-type metric perturbations couple to scalar-type photon perturbations.
We label these combinations by $\mathsf P=\mathsf S,\mathsf V$, according to whether the metric fluctuation is scalar or vector:
\begin{equation}
 \begin{aligned}
 \mathsf P=\mathsf S:\quad&
 a^{(\mathsf V)};\ \ \ \ \ \,\ h_{ab}^{(\mathsf S)},h_{a}^{(\mathsf S)},h_{1}^{(\mathsf S)},h_{2}^{(\mathsf S)},
 &&\ \ \ \ 1+7\ \text{amplitudes},\\
 \mathsf P=\mathsf V:\quad&
 a_{a}^{(\mathsf S)},a^{(\mathsf S)};\ h_{a}^{(\mathsf V)},h^{(\mathsf V)},
 &&\ \ \ \ 3+3\ \text{amplitudes}.
 \end{aligned}
 \label{eq:coupled-angular-blocks}
\end{equation}
For $\ell=0,1$ and $n=0$, some harmonics vanish and the number of active amplitudes is reduced.
Appendix~\ref{app:active-radial-components} gives the complete component counts, including these special cases.

The two blocks have opposite eigenvalues under the symmetry combining angular parity with Maxwell charge conjugation, $Q\mapsto-Q$.
Each transformation reverses the background magnetic field, so their combination preserves the saddle.
The combined symmetry has eigenvalues $(-1)^\ell$ and $(-1)^{\ell+1}$ in the $\mathsf S$ and $\mathsf V$ blocks, respectively, and invariance of the quadratic action therefore forbids mixing between them.
Here scalar and vector refer to the harmonic construction of the metric, rather than inequivalent $SO(3)$ representations at fixed $\ell$.

The magnetic saddle also preserves the discrete time-reflection symmetry (TRS) $\tau\mapsto-\tau$.
We denote its eigenvalue on the full tensor or one-form by $\eta=\pm1$.
Let $\omega_n=2\pi n/\beta$ be the Euclidean frequency.
We expand the time dependence in the real basis $\cos(\omega_n\tau)$ and $\sin(\omega_n\tau)$.\footnote{For $\eta=+1$ (TRS even), components with an even number of $\tau$ indices multiply $\cos(\omega_n\tau)$, and those with an odd number multiply $\sin(\omega_n\tau)$. For $\eta=-1$ (TRS odd), the assignments are reversed.}
After these time factors are extracted, the amplitudes are real functions of $r$.
For $n\geq1$, the time translation $\tau\mapsto\tau+\beta/(4n)$ interchanges the two TRS sectors, up to signs of radial amplitudes, so their radial spectra agree.

The final radial problems are labelled by $(n,\ell,\eta,\mathsf P)$, with $n\geq0$, $\eta=\pm1$, and $\mathsf P=\mathsf S,\mathsf V$.
Their eigenvalues have angular multiplicity $2\ell+1$.

\paragraph{Numerical methods.}
For each $(n,\ell,\eta,\mathsf P)$, substituting \eqref{eq:full-angular-decomposition} and the real time harmonics into $I^{(2)}$ gives a coupled radial eigenvalue problem on $r_h\leq r\leq r_c$.
We require the metric and gauge-field fluctuations to be smooth at both horizons, which fixes the boundary conditions for the radial differential equations.
We absorb the radial weights of \eqref{eq:coupled-ultralocal-inner-product} into the amplitudes, so that their radial norm uses the measure $\dd r$, apart from the time and angular normalization factors.
We solve the equations using Chebyshev--Gauss--Lobatto (CGL) collocation~\cite{Trefethen2000Spectral, Kolanowski:2024zrq}: the radial direction is discretized on a grid, and the differential operator is represented by a matrix acting on the field values at the grid points.
The spectrum then follows from a finite matrix eigenvalue problem.
Details of the method are described in Appendix~C.3 of Ref.~\cite{Chen:2025jqm}.
The comparison of low-lying eigenvalues and radial eigenfunctions with analytic de Sitter modes is illustrated in Subsection~\ref{sec:localized-modes}.

\subsection{Numerical spectrum}
\label{sec:spectrum-phase}
\label{sec:localized-modes}
The numerical spectrum contains both positive and negative eigenvalues.
Their signs determine the phase of the Gaussian integrals: each negative real mode contributes $-\mathrm{i}$~\cite{Maldacena:2024spf}.
The metric trace $h$, which describes the conformal factor, has a negative kinetic term and therefore gives infinitely many negative modes.
Its universal phase is absorbed into the ultralocal measure, leaving a finite excess or deficit of negative modes that determines the one-loop phase.

We denote this regularized count by $n_-$.
In each radial sector containing the metric trace $h$, we subtract one trace direction per grid point from the numerical count.\footnote{This means that at least one diagonal metric amplitude contributing to $h$ is present. Its radial amplitude is $f^{-1}h_{\tau\tau}^{(\mathsf S)}+f h_{rr}^{(\mathsf S)}+r^{-2}h_{2}^{(\mathsf S)}$. Table~\ref{tab:active-radial-components} in Appendix~\ref{app:active-radial-components} shows that it is present in the $\mathsf S$ blocks with $n\geq1$ and in the $n=0$, $\eta=+1$ $\mathsf S$ blocks, but absent from all $\mathsf V$ blocks and the $n=0$, $\eta=-1$ $\mathsf S$ blocks.}
We implement this subtraction by the prescription~\cite{Chen:2025jqm}
\begin{equation}
 n_-=
 \begin{cases}
  \text{(number of negative eigenvalues)}-n_{\rm grid},
    &\text{metric trace present},\\
  \text{(number of negative eigenvalues)},
    &\text{metric trace absent},
 \end{cases}
 \label{eq:regularized-negative-mode-count}
\end{equation}
where $n_{\rm grid}$ is the number of radial grid points.
For each angular harmonic, the remaining phase is $(-\mathrm{i})^{n_-}$.
We take the absolute value of the Faddeev--Popov determinant, interpreted as the Jacobian of the change of variables used in gauge fixing, so it contributes no phase~\cite{Polchinski:1988ua,Maldacena:2024spf}.

Along the open lukewarm family, $0<\mathsf M<1/4$, we find $n_-=-1$ in three $\ell=0$ sectors: the static sector and the two time-reflection sectors at $n=1$.\footnote{More precisely, these three sectors are $(n,\ell,\eta,\mathsf P)=(0,0,+1,\mathsf S)$ and $(1,0,\pm1,\mathsf S)$.}
Throughout this paper, ``static'' means $n=0$.
All other sectors have zero regularized count.
Their sum gives
\begin{equation}
 n_-^{\rm total}=-3,
 \qquad
 \frac{Z_{\rm bh}}{|Z_{\rm bh}|}=(-\mathrm{i})^{-3}=-\mathrm{i}.
 \label{eq:bh-one-loop-phase}
\end{equation}

To interpret this result, we first compare the small-mass spectrum sector by sector with empty de Sitter and a probe observer, and then follow it along the lukewarm family.
Let $\ell_0$ label scalar harmonics on $S^4$, while $\ell$ labels harmonics on $S^2$.
The constant $\ell_0=0$ mode and the five $\ell_0=1$ conformal modes are absent from the negative conformal spectrum.
To determine how the observer changes this count, we ask which modes leave its worldline fixed and which move it.
The constant mode has $n=\ell=0$.
The five $\ell_0=1$ modes, whose gradients generate proper conformal Killing vectors, split as
\begin{equation}
 \ell_0=1
 \quad\longrightarrow\quad
 (n,\ell,\eta)=(1,0,\pm1)\oplus(0,1,+1).
 \label{eq:ckv-sector-decomposition}
\end{equation}
The two $\ell=0$ modes leave the observer's worldline fixed, while the three $\ell=1$ modes move it in the three spatial directions.

The three missing $\ell=1$ modes contribute a phase $(-\mathrm{i})^{-3}$.
The observer has three negative static displacement modes, contributing $(-\mathrm{i})^3$ and canceling this factor.
The constant mode and the two $\ell=0$ modes therefore leave precisely the three deficits found in the black hole calculation.
In particular, the black hole spectrum has zero regularized count at $n=0$, $\ell=1$, and retains the three deficits at $\ell=0$.
It therefore gives the same phase as the probe-observer calculation, with the black hole backreaction included.
We identify the corresponding localized black hole modes below and in Section~\ref{sec:translation-worldline}.

We can then follow the count as the mass changes along the lukewarm family.
A continuous change of mass can change the count only if an eigenvalue crosses zero.
We find no such crossing or exact zero mode in the coupled bosonic spectrum for $0<\mathsf M<1/4$.
The phase therefore remains $-\mathrm{i}$, indicating that the whole lukewarm line can be interpreted as a finitely backreacting observer.

\begin{figure}[t]
 \centering
 \includegraphics[width=\textwidth]{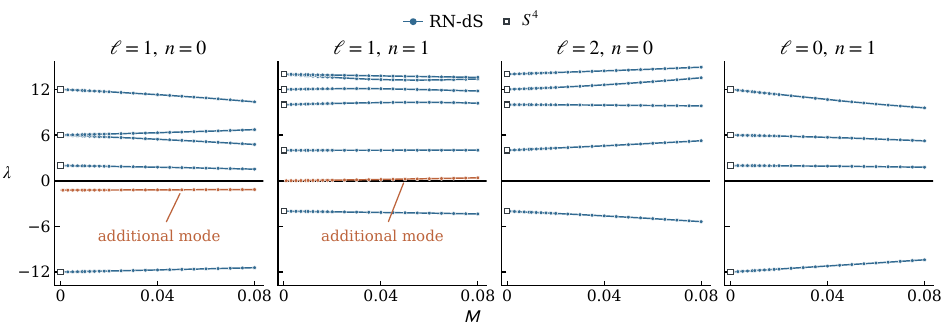}
 \caption{Illustration of eigenvalues in different sectors as $\mathsf M\rightarrow0$.
 Squares mark the corresponding de Sitter eigenvalues; dots are numerical results, solid curves interpolate between them, and dashed curves are small-mass fits extrapolated to $\mathsf M=0$.
 The solid black line marks $\lambda=0$. All panels show $\eta=+1$ in the $\mathsf P=\mathsf S$ block. The ``additional'' modes only appear in the $\ell=1$ panels.}
 \label{fig:small-mass-spectrum}
\end{figure}

\paragraph{The spectrum in the small-mass limit.}
Having understood the phase, we now examine the spectrum as the black hole shrinks.
At fixed nonzero $r/l_\text{dS}$, the background approaches empty de Sitter as $\mathsf M\to0$, so one expects much of the spectrum to approach the sphere spectrum.
Figure~\ref{fig:small-mass-spectrum} illustrates this behavior: most eigenvalues shown approach the corresponding integer de Sitter values.\footnote{The sphere harmonics and their Laplacian identities are collected in Ref.~\cite{Law:2020cpj}. In the normalization of \eqref{eq:dimensionless-fluctuation-operator}, the trace, scalar-derived traceless, vector-derived, and transverse-traceless metric modes on $S^4$ have eigenvalues $6-\ell_0(\ell_0+3)$, $\ell_0(\ell_0+3)-6$, $\ell_0(\ell_0+3)-4$, and $\ell_0(\ell_0+3)$, respectively; transverse Maxwell modes have eigenvalues $\ell_0(\ell_0+3)+2$. Here $\ell_0$ denotes the $S^4$ harmonic degree, and only harmonics allowed in the given sector are included.}
In the low-lying $\ell=1$ spectrum, however, each  $n$ also has a mode whose limiting eigenvalue does not match a de Sitter fluctuation in the corresponding sector.
We distinguish these additional modes by examining their eigenfunctions.

\begin{figure}[t]
 \centering
 \includegraphics[width=0.8\textwidth]{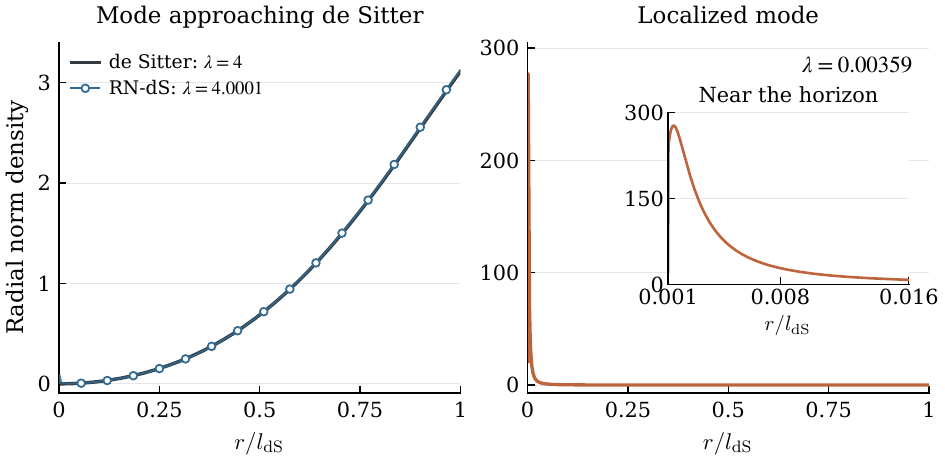}
 \caption{Comparison of the norm densities of two eigenmodes in the $\ell=1,\,n=1,\,\eta=+1, \mathsf P=\mathsf S$ sector at $\mathsf M=0.001$.
 The left panel compares the eigenfunction of an extended mode (from numerical diagonalization) with its analytic de Sitter counterpart; the right panel shows the localized mode.
 }
 \label{fig:ds-localized-wavefunctions}
\end{figure}

The radial norm densities show how these modes differ (Fig.~\ref{fig:ds-localized-wavefunctions}).
The mode whose eigenvalue approaches a de Sitter value extends over the radius $l_\text{dS}$, and its norm density nearly coincides with that of the corresponding analytic de Sitter mode.
The additional mode is instead concentrated at radii $r=O(r_h)$, approaching the center $r=0$ of the limiting de Sitter static patch as $\mathsf M\to0$.
Its radial width decreases with the black hole horizon radius.
The mass dependence is illustrated in Fig.~\ref{fig:translation-squeezing} in Section~\ref{sec:translation-worldline}.

These localized modes suggest a description in terms of the observer's motion: a small black hole moving like a heavy particle in de Sitter.
The central Euclidean worldline is the great circle at $r=0$, with period $2\pi l_\text{dS}$.
For a particle in de Sitter, the fluctuation operator for a small transverse displacement about this worldline is
$-\partial_\tau^2-l_\text{dS}^{-2}$.
Its Fourier eigenfunctions have eigenvalues $(n^2-1)/l_\text{dS}^2$, or $n^2-1$ in de Sitter units, as in the observer worldline calculation of Ref.~\cite{Maldacena:2024spf}.

A small displacement $\boldsymbol Z(\tau)\in\mathbb R^3$ gives the radial shift $\delta r=\boldsymbol Z\cdot\widehat{\boldsymbol r}$, where $\widehat{\boldsymbol r}\in\mathbb R^3$ is the unit radial vector.
The components of $\widehat{\boldsymbol r}$ are $\ell=1$ harmonics, so a displacement produces $\ell=1$ perturbations, matching the localized modes.

\begin{figure}[t]
 \centering
\includegraphics[width=\textwidth]{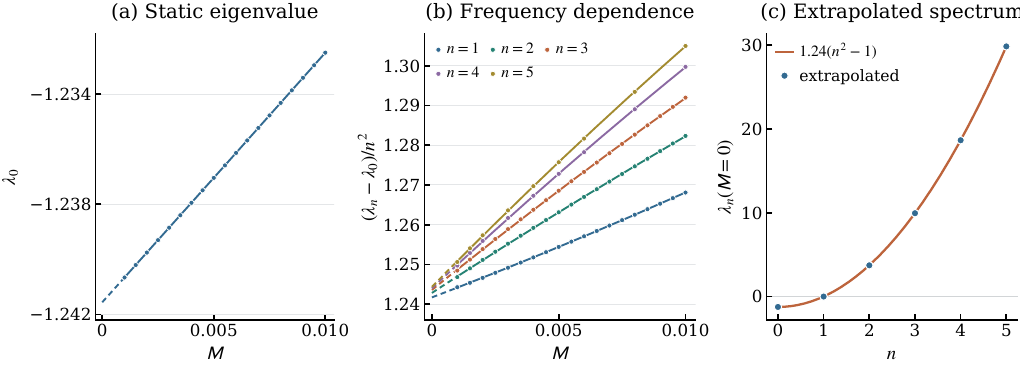}
 \caption{The small-mass limits agree with the $n^2-1$ dependence predicted by the particle action.
 Dashed curves show fits used to estimate these limits.
 }
 \label{fig:localized-small-mass-fits}
\end{figure}

Let $\lambda_n$ denote the eigenvalue of the localized mode with Fourier number $n$ and angular momentum $\ell=1$.
The numerical small-mass limits support the worldline interpretation (Fig.~\ref{fig:localized-small-mass-fits}):
\begin{equation}
 \lambda_n\simeq1.24\,(n^2-1),
 \qquad {n\geq 0}, \qquad \mathsf M\to0.
 \label{eq:localized-spectrum-fit}
\end{equation}

Later in section~\ref{sec:translation-worldline}, we explain the origin of this family by relating it to the translation zero mode of an extremal RN black hole in asymptotically flat spacetime and calculating its lift by de Sitter curvature.

\paragraph{Further comment on the phase.}
The worldline picture also provides a check of the phase interpretation above.
If the localized modes have the limiting spectrum in \eqref{eq:localized-spectrum-fit}, their three static modes are negative, as required to cancel the three missing conformal negative modes in the $\ell=1$ sector of the graviton spectrum on empty $S^4$.
For each fixed $n\geq2$, the localized eigenvalues are positive at sufficiently small mass.
The only remaining sign to determine is therefore at $n=1$, where the leading worldline eigenvalue vanishes.

For a small black hole, we approximate the $n=1$ mode by the metric and Maxwell perturbations generated by a displacement (vector) whose zero-mass limit is
a Killing vector on $S^4$.
Let $u_\xi$ denote the resulting (metric and Maxwell) perturbation.
The Einstein--Maxwell action is unchanged by the transformation, so only the gauge-fixing term, which is a square, contributes
\begin{equation}
 I^{(2)}[u_\xi]=I_{\rm gf}^{(2)}[u_\xi]\geq0.
 \label{eq:n1-displacement-positive-action}
\end{equation}
Appendix~\ref{app:nonstatic-matching} constructs this displacement by matching and finds a positive gauge-fixing action for small, nonzero mass.
Since the norm is positive, the displacement picture predicts that the $n=1$ eigenvalue approaches zero from above and adds no negative modes.
This agrees with the positive $n=1$ eigenvalues in Fig.~\ref{fig:small-mass-spectrum} and supports the phase count above.

Section~\ref{sec:translation-worldline} constructs the corresponding metric and Maxwell perturbations, explains the coefficient $1.24$ in \eqref{eq:localized-spectrum-fit}, and derives their worldline action in the small-mass limit.

\section{Observer worldline action from localized modes}
\label{sec:translation-worldline}
In this section, we explain the localized modes in the small-mass limit and derive their worldline action from the translation zero mode of an extremal Reissner--Nordstr\"om (RN) black hole in asymptotically flat spacetime.
De Sitter curvature and the associated finite-temperature horizon lift this zero mode; time dependent displacement gives the kinetic term in the worldline action.

The horizon radius is of order $O(l_\text{dS}\mathsf M)$, as follows from \eqref{eq:lukewarm-parameters}.
For perturbations varying on this scale, the dimensionless Lichnerowicz operator generically has eigenvalues of order $O(\mathsf M^{-2})$.
The localized eigenvalues instead remain $O(1)$, indicating a zero mode of the leading operator.

To identify it, we measure radius in black hole units, $x=r/(l_\text{dS}\mathsf M)$.
As $\mathsf M\to0$ at fixed $x>1$, the blackening factor tends to that of an extremal RN black hole in asymptotically flat spacetime, which we call flat-RN.\footnote{The overlap $\mathsf M\ll r/l_\text{dS}\ll1$ lies far outside the black hole horizon and well inside the cosmological horizon. Both the black hole and de Sitter corrections to the flat metric are small there.}
Let $\mathcal O_0$ denote the dimensionless Lichnerowicz operator in the static sector.
In black hole units, $\mathsf M^2\mathcal O_0\to\mathcal L_0$.
Write the radial eigenfunction as $q(x)\to q_0(x)\ne0$.\footnote{The vector $q$ contains weighted radial amplitudes of $u=(h_{\mu\nu},a_\mu)$. After integrating the ultralocal inner product over time and angles, we absorb the radial weights from the volume element and tensor contractions into these amplitudes, so that the remaining norm has a flat radial measure $\dd r$. Appendix~\ref{app:regions-translation-data} gives the five-component vector and its normalization in the static $\ell=1$ sector.}
Since the numerical eigenvalue remains $O(1)$, the eigenvalue equation gives
\begin{equation}
 \mathsf M^2\mathcal O_0q=\mathsf M^2\lambda q
 \quad\xrightarrow{\mathsf M\to0}\quad
 \mathcal L_0q_0=0.
 \label{eq:translation-kernel}
\end{equation}
Thus the leading localized eigenfunction is a flat-RN zero mode.\footnote{For related studies of modes localized near the horizon in the full Euclidean fluctuation spectrum of near-extremal black holes, see Ref.~\cite{Kolanowski:2024zrq}. The modes considered there are associated with Schwarzian, $U(1)$ gauge, and rotational fluctuations.}

A static black hole in asymptotically flat spacetime can be translated without changing its action.
The three Cartesian translations give zero modes in the $n=0$, $\ell=1$ sector.
This statement does not require extremality: Gross, Perry, and Yaffe constructed the corresponding modes for Schwarzschild~\cite{Gross:1982cv}.
In Subsection~\ref{sec:translation-identification}, we construct the charged extremal translation and compare it with the static localized eigenfunction.

For the asymptotically flat RN family, let $\beta_{\rm RN}$ denote the inverse Hawking temperature.
In the extremal limit, $\beta_{\rm RN}^{-1}\to0$.
On a Fourier harmonic $e^{2\pi \mathrm{i} n\tau/\beta_{\rm RN}}$ with fixed $n$, the time derivative acts as $\partial_\tau=2\pi \mathrm{i} n/\beta_{\rm RN}\to0$.
Thus the same radial translation zero mode also governs the fixed-$n$ limit for $n>0$.

Returning to the lukewarm family, the physical period tends to $2\pi l_\text{dS}$, while the period in black hole units, $\beta/(l_\text{dS}\mathsf M)$, diverges as $\mathsf M\to0$.
Consequently, $l_\text{dS}\mathsf M\omega_n\to0$ for each fixed $n$, and the leading radial equation at fixed $x>1$ is again the static flat-RN equation.
De Sitter curvature, the smooth finite-temperature cap, and the subleading time-derivative terms lift this limiting zero mode.

We describe this family by a worldline effective field theory~\cite{Goldberger:2004jt,Kol:2007rx}, with the black hole displacement $\boldsymbol Z(\tau)$ as its collective variable.\footnote{Worldline effective theories for electrically charged black holes in a spacetime with a compact spatial direction, including their extremal limit, were developed in Ref.~\cite{Gilmore:2009ea}.}
The localized eigenvalues approach $\lambda_n\simeq1.24(n^2-1)$.
We match the quadratic action $\lambda_n\|u_n\|_{\rm ultra}^2/(2G_Nl_\text{dS}^2)$ of these eigenmodes to the de Sitter particle action~\cite{Maldacena:2024spf}.
Here $u_n$ denotes the coupled metric and Maxwell perturbations, and a dot denotes $\dd/\dd\tau$:
\begin{equation}
 I^{(2)}_{\rm wl}[\boldsymbol Z]
 =\frac M2\int_0^{2\pi l_\text{dS}}\dd\tau
 \left(\dot{\boldsymbol Z}^{\,2}
             -\frac{\boldsymbol Z^{\,2}}{l_\text{dS}^2}\right).
 \label{eq:probe-worldline-comparison}
\end{equation}
Subsection~\ref{sec:static-translation-lifting} computes the static lift by perturbing the extremal flat-RN translation and imposing smoothness at the finite-temperature horizon.
It gives $\lambda_0\simeq-1.24$, in agreement with the independent numerical diagonalization.
Thermodynamically, this lift reflects the change in action when the black hole is displaced into a hotter local bath.
Subsection~\ref{sec:translation-worldline-matching} then includes time dependence and recovers the observed $n^2-1$ dependence.

The normalization is essential for converting the spectrum into an action for $\boldsymbol Z(\tau)$.
At a fixed displacement $\boldsymbol Z(\tau)$ of order $O(1)$, the weighted radial eigenfunction has $O(1)$ amplitude and occupies a radial interval of width $O(l_\text{dS}\mathsf M)$.
The ultralocal normalization integral $\|u\|_{\rm ultra}^2$ therefore scales as $O(\mathsf M)$ at fixed $l_\text{dS}$.
Together with the $O(1)$ eigenvalue, this gives an $O(\mathsf M)$ quadratic action, as required by the worldline action for a mass $M=l_\text{dS}\mathsf M/G_N$; Subsection~\ref{sec:translation-worldline-matching} gives the detailed calculation.

\subsection{The translation zero mode}
\label{sec:translation-identification}
In this subsection, we work in the extremal RN geometry in asymptotically flat spacetime, construct its static translation zero mode, and compare it with the small-mass limit of the $n=0$ localized RN--dS eigenmode in Subsection~\ref{sec:localized-modes}.

\paragraph{Translation in asymptotically flat space.}
A black hole in asymptotically flat spacetime can be translated relative to a fixed asymptotic frame without changing its action.
Following the construction of Gross, Perry, and Yaffe (GPY) for Schwarzschild in de Donder gauge~\cite{Gross:1982cv}, we construct this mode for the charged extremal RN black hole.
The background metric is
\begin{equation}
 \dd\bar s^2=f_{\rm RN}(r)\dd\tau^2+f_{\rm RN}(r)^{-1}
 {\dd r^2}+r^2\dd\Omega_2^2.
 \label{eq:translation-rn-background}
\end{equation}
Here $f_{\rm RN}(r)=\left(1-\frac{\mathsf{M}l_\text{dS}}{r}\right)^2$ and the magnetic field is $\bar F=Q\sin\theta\,\dd\theta\wedge\dd\varphi$.

An infinitesimal translation is generated by the coordinate change $x^\mu\mapsto x^\mu-\xi^\mu(x)$.
We keep the asymptotic frame fixed, so the value of $\xi$ at infinity specifies the displacement.
By spherical symmetry, we may choose the $z$ direction and require $\xi\to Z(\tau)\partial_z$, where $Z(\tau)$ is the displacement amplitude and $\partial_z=\cos\theta\,\partial_r-(\sin\theta/r)\partial_\theta$.
The corresponding metric and Maxwell-potential perturbations are
\begin{equation}
 h_{\mu\nu}=2\nabla_{(\mu}\xi_{\nu)},
 \qquad
 a_\mu=\xi^\nu\bar F_{\nu\mu},
 \label{eq:translation-coupled-fields}
\end{equation}
where $\nabla$ is the flat-RN covariant derivative.
For a static translation, these fields solve the linearized Einstein--Maxwell equations because they are generated from an exact solution.\footnote{Calling the perturbation a zero mode also requires the appropriate boundary conditions: we impose horizon regularity and falloff of the metric and Maxwell perturbations at infinity.}
Their radial form depends on the gauge.
To compare with the numerical eigenmodes, we impose the de Donder and Lorenz conditions used in Section~\ref{sec:one-loop},
\begin{equation}
 \nabla^2\xi_\mu+R_{\mu\nu}\xi^\nu=0,
 \qquad \nabla^\mu a_\mu=0,
 \label{eq:translation-gauge-conditions}
\end{equation}
where $R_{\mu\nu}$ is the Ricci tensor.
Write $\xi$ in terms of two dimensionless radial functions $P(x)$ and $W(x)$:
\begin{equation}
 \xi=Z(\tau)\left[P(x)\cos\theta\,\partial_r
       -\frac{W(x)}{r}\sin\theta\,\partial_\theta\right].
 \label{eq:translation-vector-problem}
\end{equation}
Constant $Z$ gives the static translation.
For a harmonic $Z(\tau)\propto e^{2\pi \mathrm{i} n\tau/\beta_{\rm RN}}$, we have $\partial_\tau Z=(2\pi \mathrm{i} n/\beta_{\rm RN})Z\to0$ at fixed $n$ as $\beta_{\rm RN}^{-1}\to0$. The leading radial functions $P,W$ therefore also apply to the fixed-$n$ zero-temperature limit. On the lukewarm family, the corresponding statement is $l_\text{dS}\mathsf M\omega_n\to0$ as $\mathsf M\to0$.
The frequency corrections are treated in Subsection~\ref{sec:translation-worldline-matching}.
A unit translation at large $x$ requires
\begin{equation}
 P(x),W(x)\longrightarrow1\qquad(x\to+\infty).
 \label{eq:translation-unit-displacement}
\end{equation}
Substituting the static ansatz into the de Donder condition gives two linear radial equations.
We solve them with this asymptotic condition and the horizon condition\footnote{The horizon smoothness condition is defined by taking the extremal limit of a smooth near-extremal solution.} specified in Appendix~\ref{app:regions-translation-data}, then reconstruct $h_{\mu\nu}$ and $a_\mu$ using \eqref{eq:translation-coupled-fields}.

\begin{figure}[t]
 \centering
 \includegraphics[width=0.88\textwidth]{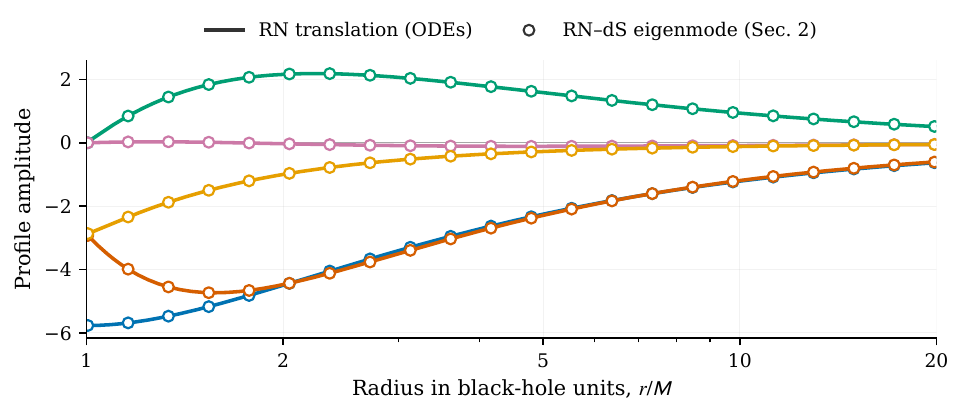}
 \caption{Comparison of the flat-RN translation with the static ($n=0$) localized RN--dS eigenmode at $\mathsf M=0.005$.
 Solid curves show the flat-RN translation obtained from its radial equations; open circles show the RN--dS eigenmode from the independent diagonalization in Section~\ref{sec:one-loop} (different color for different components).
 Both eigenfunctions are expressed in the same weighted combinations of metric and Maxwell components, defined in Appendix~\ref{app:regions-translation-data} in detail.}
 \label{fig:translation-comparison}
\end{figure}

\paragraph{Comparison with the localized mode.}
The flat-RN translation and the static ($n=0$) localized RN--dS eigenmode are obtained by independent calculations in the same gauge.
To compare their eigenfunctions as the black hole shrinks, we use $x=r/(l_\text{dS}\mathsf M)$.
At fixed $x>1$, the de Sitter correction $-\mathsf M^2x^2$ vanishes, leaving the extremal flat-RN geometry.
The approximately flat interval $\mathsf M\ll r/l_\text{dS}\ll1$, far outside the black hole horizon and well inside the cosmological horizon, provides the reference frame for the displacement.
Figure~\ref{fig:translation-comparison} shows that all five metric and Maxwell amplitudes agree closely after a single overall rescaling, supporting the identification of the localized mode with a black hole translation.

We next calculate how de Sitter curvature lifts the translation eigenvalue from zero.

\subsection{Lifting the translational zero mode by de Sitter curvature}
\label{sec:static-translation-lifting}
In this subsection, we restrict to static displacements.
A thermodynamic argument first determines the expected change in action: a small lukewarm black hole is in equilibrium with the surrounding bath, and displacing it outward in de Sitter places it in a hotter local bath.

At leading order in $\mathsf M$, we use the de Sitter geometry to describe this bath.
The unperturbed black hole is represented by a worldline at the center of the static patch, $r=0$; its displacement is $\boldsymbol Z$.
The local inverse temperature obeys the Tolman relation~\cite{Tolman:1930ona} and equals the proper circumference of the Euclidean time circle,
\begin{equation}
 \beta_{\mathrm{loc}}=\beta \sqrt{1-\frac{\boldsymbol Z^2}{l_\text{dS}^2}}.
 \label{eq:static-local-temperature}
\end{equation}
An outward displacement decreases this circumference.

At fixed magnetic charge, the black hole contribution to the action is $I_{\mathrm{th}}=\beta_{\mathrm{loc}} E-S$~\cite{Gibbons:1976ue,Mann:1995vb}, where $E$ and $S$ are its energy and entropy.
Its variation is
\begin{equation}
    \dd I_{\mathrm{th}}
 =E\,\dd\beta_{\mathrm{loc}}
   +\beta_{\rm loc}\dd E -\dd S =  E\,\dd\beta_{\mathrm{loc}}
   +(\beta_{\rm loc}-\beta_H)\dd E,
 \label{eq:static-first-law}
\end{equation}
where we used the fixed-charge first law $\dd S=\beta_H\dd E$, with $\beta_H$ the inverse Hawking temperature.
In local thermal equilibrium, $\beta_{\rm loc}=\beta_H$, so $\dd I_{\mathrm{th}}=E\dd\beta_{\rm loc}$.
Expanding \eqref{eq:static-local-temperature} for small $\boldsymbol Z$ gives
\begin{equation}
 \Delta I_{\mathrm{th}}
 =-\frac{\beta E}{2l_\text{dS}^2}\boldsymbol Z^2, \quad
 \text{with} \quad E=\frac{l_\text{dS}\mathsf M}{G_N}[1+O(\mathsf M)].
 \label{eq:static-thermal-prediction}
\end{equation}
This agrees with the static term in \eqref{eq:probe-worldline-comparison}, which we now derive from the lift of the translation zero mode.

\paragraph{Lifting the translation zero mode.}
De Sitter curvature corrects both the flat-RN Lichnerowicz operator and its translation eigenfunction.
To determine the leading eigenvalue, we expand $\mathsf M^2\mathcal O=\mathcal L_0+\mathsf M^2\mathcal L_2+\cdots$ and $q(x)=q_0(x)+\mathsf M^2q_2(x)+\cdots$.
Here $\mathcal L_0$ is the extremal flat-RN operator, $q_0$ is its translation zero mode, and $\mathcal L_2$ is the de Sitter curvature correction, given explicitly in Appendix~\ref{app:static-time-matching}.
The eigenvalue equation is\footnote{The static bulk operator has no term linear in $\mathsf M$. An eigenfunction correction at that order obeys the homogeneous zero-mode equation. Assuming that no other static zero mode mixes with the translation, fixing the displacement removes this correction.}
\begin{equation}
    \mathcal{O} q(x)= \lambda_0 q(x),
    \label{eq:static-full-eigenvalue-equation}
\end{equation}
Its leading order gives $\mathcal L_0q_0=0$, while the next order gives
\begin{equation}
     \mathcal L_0q_2+\mathcal L_2q_0=\lambda_0(\mathsf M = 0)q_0\,.
     \label{eq:static-response-equation}
\end{equation}
Here $\lambda_0(\mathsf M=0)$ denotes the limiting eigenvalue to be determined in the following.

We do not need to solve for $q_2$ throughout the geometry.
Taking the inner product with $q_0$ and integrating by parts moves $\mathcal L_0$ onto $q_0$.
Since $\mathcal L_0q_0=0$, the contribution of $q_2$ reduces to a boundary term.
The radial inner product is
\begin{equation}
 \langle u,v\rangle=\int_1^\infty\dd x\,u^{\mathsf T}v,
 \label{eq:static-radial-norm}
\end{equation}
for real five-component radial fields $u,v$.
The limits are the horizon and spatial infinity of the limiting flat-RN geometry, and the normalization is fixed by $P(x),W(x)\to1$ in \eqref{eq:translation-unit-displacement}.
Subsection~\ref{sec:translation-worldline-matching} restores the measure $\dd r=l_\text{dS}\mathsf M\dd x$ and the time and angular integrals to obtain the physical ultralocal normalization at fixed displacement.
{We denote the radial norm of this translation by $\mathcal N_0=\langle q_0,q_0\rangle$.}
The projection therefore gives
\begin{equation}
    \frac{\langle q_0(x),\mathcal{L}_2 q_0(x)\rangle+ \text{``boundary term''}}{\mathcal N_0} = \lambda_0(\mathsf M = 0).
    \label{eq:static-curvature-projection}
\end{equation}
The bulk matrix element is determined by the flat-RN translation.
The boundary term would vanish with the extremal flat-RN horizon condition, but RN--dS requires a smooth horizon at finite temperature.
We must include the correction associated with this change of horizon condition.

\begin{figure}[t]
 \centering
 \includegraphics[width=\textwidth]{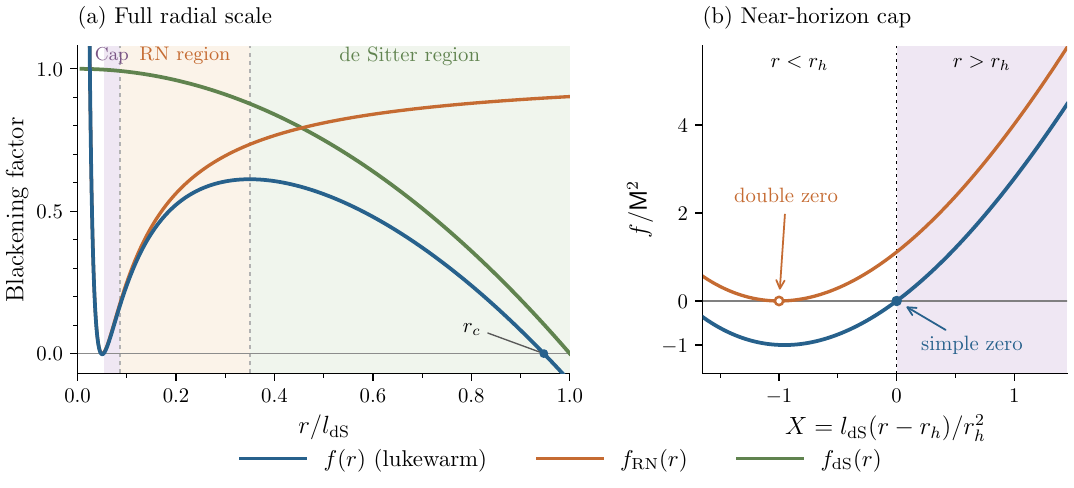}
 \caption{Exact blackening factors $f(r)$, $f_{\rm RN}(r)=(1-l_\text{dS}\mathsf M/r)^2$, and $f_{\rm dS}(r)=1-r^2/l_\text{dS}^2$, at $\mathsf M=0.05$.
 The bands and dividers indicate the three regions schematically; the left cap band is widened for visibility.
 The right panel magnifies the near-horizon region using $X=l_\text{dS}(r-r_h)/r_h^2$ and $f/\mathsf M^2$.
 The RN factor has a double zero at $r=l_\text{dS}\mathsf M$, exactly $X=-1$, whereas the lukewarm factor has a simple outer zero at $r=r_h$, or $X=0$.
 Their separation is $r_h-l_\text{dS}\mathsf M=r_h^2/l_\text{dS}=O(l_\text{dS}\mathsf M^2)$.}
 \label{fig:expansion-regions}
\end{figure}

\paragraph{Horizon smoothness.}
The flat-RN limit has zero Hawking temperature, whereas the lukewarm black hole has finite temperature.
Reconciling these descriptions requires resolving the near-horizon cap, where the extremal approximation fails.\footnote{For a related matched-asymptotic treatment of coupled gravitational and electromagnetic perturbations of near-extremal RN, see Ref.~\cite{Porfyriadis:2018jlw}, which considers Lorentzian, asymptotically flat electric backgrounds and even-parity modes with $\ell\geq2$.}
Near $x=1$, the metric function and horizon position obey
\begin{equation}
\begin{aligned}
f(x)&=\left(1-\frac1x\right)^2-\mathsf M^2x^2
\simeq(x-1)^2-\mathsf M^2
\qquad(x-1\sim O(\mathsf M)),\\
x_h&\equiv\frac{r_h}{l_\text{dS}\mathsf M}
=1+\mathsf M+O(\mathsf M^2).
\end{aligned}
\label{eq:static-horizon-scale}
\end{equation}
The curvature term $-\mathsf M^2x^2$ becomes comparable to $(1-1/x)^2$ when $x-1\sim O(\mathsf M)$.
This cap region, $r-l_\text{dS}\mathsf M\sim O(l_\text{dS}\mathsf M^2)$, has radial coordinate width $O(l_\text{dS}\mathsf M^2)$.
Outside the cap but still at $r\ll l_\text{dS}$, the geometry is approximated by flat-RN; at $r\sim O(l_\text{dS})$, the de Sitter curvature must be retained.
Figure~\ref{fig:expansion-regions} illustrates these three regions and their blackening factors.

Despite its small coordinate width, the cap can contribute to the leading eigenvalue.
Since $f\sim O(\mathsf M^2)$ there, its proper radial size is $O(l_\text{dS}\mathsf M)$, and the Euclidean time circle contracts from circumference $O(\beta\mathsf M)$ to zero.
Resolving the coordinate width gives $\partial_r\sim O((l_\text{dS}\mathsf M^2)^{-1})$, so the radial derivative terms scale as
\begin{equation}
    \mathsf M^2\mathcal O \sim \mathsf M^2 l_\text{dS}^2f\,\partial_r^2\sim O(1).
    \label{eq:static-cap-power-counting}
\end{equation}
Thus the local operator is not suppressed in the cap.
The matched action calculation below gives an $O(\mathsf M^2)$ correction to the rescaled radial action, which the overall $\mathsf M^{-2}$ in $\mathcal O$ converts into an $O(1)$ eigenvalue shift.
We therefore impose horizon smoothness in the cap before evaluating its contribution.
We use the coordinate
\begin{equation}
 X=\frac{l_\text{dS}(r-r_h)}{r_h^2}
  =\frac{x-x_h}{\mathsf M x_h^2}.
 \label{eq:static-cap-coordinate}
\end{equation}
It keeps the physical horizon at $X=0$.
Appendix~\ref{app:expansion-regions} gives the corresponding expansion.
The cap and RN solutions must agree in the overlap $\mathsf M\ll x-x_h\ll1$, or $1\ll X\ll\mathsf M^{-1}$.

To perform the matching, we choose a sphere in this overlap and equate the physical metric and Maxwell perturbations on its two sides.
Horizon smoothness selects the cap solution, while $P(x),W(x)\to1$ fixes the normalization of the flat-RN solution.
The action is the sum of the contributions from the adjoining regions, with their matching-surface terms combined.
Because the flat-RN calculation already includes the extremal throat, we must subtract that throat contribution when replacing it with the smooth finite-temperature cap.
We denote the resulting correction in \eqref{eq:static-curvature-projection} by $\mathcal J_2$.

Its magnitude is controlled by the near-horizon amplitude of the translation normalized at infinity.
The radial equations and the conditions at both ends give\footnote{The coefficient $a_0$ is fixed by the connection matrix for the radial equations, as in \eqref{eq:app-rn-connection-system}.}
\begin{equation}
    W(x)\to a_0,\qquad
P(x)\simeq2a_0(x-1)^2
\qquad(x\to1^+).
\label{eq:static-translation-horizon-profile}
\end{equation}
Matching fixes the leading cap solution to have the same amplitude $a_0$.
Since the action is quadratic, the correction is proportional to $a_0^2$.
Comparing the cap and extremal-throat actions, including the integration-by-parts conversion to the operator convention used for $\mathcal L_2$, gives
\begin{equation}
\mathcal J_2=a_0^2.
\label{eq:static-horizon-response}
\end{equation}
Appendix~\ref{app:static-time-matching} gives the matched solution and the action calculation.
Adding this correction to the curvature matrix element gives\footnote{We assume no additional contribution at this order from the de Sitter region $r=O(l_\text{dS})$.}
\begin{equation}
\lambda_0
=\frac{\langle q_0,\mathcal L_2q_0\rangle+a_0^2}
{\mathcal N_0}+o(1) \approx -1.24156.
\label{eq:static-completed-projection}
\end{equation}
This agrees with the independently computed numerical eigenvalue in Subsection~\ref{sec:localized-modes}.

Comparing this eigenvalue with the thermal prediction in \eqref{eq:static-thermal-prediction} also requires the normalization at fixed displacement.
In the next subsection, we compute the lift of the localized mode in the time-dependent case and this normalization factor, thereby reconstructing the kinetic term proportional to $\dot{\boldsymbol Z}^{\,2}$ in the worldline action.

\subsection{Dynamics and the worldline action}
\label{sec:translation-worldline-matching}
We now allow the black hole displacement to depend on Euclidean time and determine the kinetic term.
Expand the displacement in Fourier modes,
\begin{equation}
 \boldsymbol Z(\tau)=\sum_{n\in\mathbb Z}\boldsymbol Z_n e^{\mathrm{i}\omega_n\tau},
 \qquad \omega_n=\frac{2\pi n}{\beta},
 \qquad \boldsymbol Z_{-n}=\boldsymbol Z_n^*.
 \label{eq:worldline-displacement-frequency}
\end{equation}

Let us see how this changes the localized mode.
Naively, we can just try to multiply the static mode $q_0(x)$ by $e^{i\omega_n \tau}$.
Indeed, for the blackhole which has a characteristic time scale to be $l_\text{dS}\mathsf M$, $ \omega_n \sim n/l_\text{dS}$ is a pretty slow motion.
However, the complete description of the non-static mode is more complicated.
First, time derivatives additionally source $h_{\tau r}$ and $h_{\tau i}$ perturbations.
Second, in the cap region, the simple picture of multiplying the static mode with $e^{i\omega_n \tau}$ does not hold because we require the perturbation to be smooth at the horizon.\footnote{As a simple example, any smooth time dependent perturbation $h(r)e^{i \omega \tau}$ has to satisfy $h(r)\to 0$ as $r\to 0$, otherwise the $r\to 0$ limit is ill-defined.}
Both differences would require a refined discussion of the matching procedure used to determine the localized mode, see  Appendix~\ref{app:nonstatic-matching}.

\paragraph{The kinetic term coefficient.}
With these matched solutions, we can compute the action, in particular the kinetic term capturing the time derivatives.
For $|n|>0$, at fixed $x>1$, the seven-component Lichnerowicz operator has the schematic expansion
\begin{equation}
 \mathsf M^2\mathcal O
 =\widetilde{\mathcal L}_0
 +\mathrm{i} n\mathsf M\,\widetilde{\mathcal L}_1
 +\mathsf M^2
  \left(n^2\widetilde{\mathcal L}^{\omega}_2
        +\widetilde{\mathcal L}_2\right)
 +\cdots .
 \label{eq:worldline-operator-expansion}
\end{equation}
Here $\widetilde{\mathcal L}_0$ is the leading flat-RN operator, $\widetilde{\mathcal L}_1$ comes from the time derivative and couples the five fields in the static case to the two new fields from time derivatives, and $\widetilde{\mathcal L}_2^\omega$ is the coefficient of $\partial_\tau^2$ in the Lichnerowicz operator.
The operator $\widetilde{\mathcal L}_2$ contains the curvature correction.

As in the static calculation in Section~\ref{sec:static-translation-lifting}, we expand the eigenvalue equation about the flat-RN translation:
\begin{equation}
 \mathcal O q_n=\lambda_n q_n,
 \qquad
 q_n=q_0+\mathsf M q_{1,n}+\mathsf M^2q_{2,n}+\cdots .
\end{equation}
Here $q_0$ is extended by zeros for the two additional metric components.
The first two orders give
\begin{equation}
 \widetilde{\mathcal L}_0q_0=0,
 \qquad
 \widetilde{\mathcal L}_0q_{1,n}
 =-\mathrm{i}n\widetilde{\mathcal L}_1q_0.
\end{equation}
Thus time derivatives source a correction to the localized mode at $O(n\mathsf M)$.
At order $\mathsf M^2$, the eigenvalue equation gives
\begin{equation}
 \widetilde{\mathcal L}_0q_{2,n}
 +\mathrm{i}n\widetilde{\mathcal L}_1q_{1,n}
 +\left(n^2\widetilde{\mathcal L}_2^\omega
       +\widetilde{\mathcal L}_2\right)q_0
 =\lambda_n(\mathsf M=0)q_0.
\end{equation}
After using the above two equations and some algebra, we can express the term proportional to $n^2$ in the quadratic action as \footnote{Using the radial equations, the combined integrand reduces to a total radial derivative, allowing an analytic evaluation; see Appendix~\ref{app:nonstatic-matching}.}
\begin{equation}
 \begin{aligned}
 \mathcal{I}\supset{}&n^2\mathsf M^2
 \left\langle q_0,\,
 \left(
 \widetilde{\mathcal L}^{\omega}_2
 +\widetilde{\mathcal L}_1
  \widetilde{\mathcal L}_0^{-1}
  \widetilde{\mathcal L}_1
 \right)q_0\right\rangle = n^2\mathsf M^2\left(\frac32-\frac34a_0^2\right).
 \end{aligned}
 \label{eq:worldline-kinetic-perturbation}
\end{equation}
where $a_0$ is used in~\eqref{eq:static-translation-horizon-profile}.

Once again, this is not yet the complete action: we must also include the contribution from the region very close to the horizon.
At nonzero frequency, the RN fields decay near the horizon, and expanding at fixed $x$ misses this behavior.
Restoring it contributes $3a_0^2/4$ to the coefficient of $n^2\mathsf M^2$, so the complete frequency-dependent term is
\begin{equation}
 n^2\mathsf M^2\left[
 \left(\frac32-\frac34a_0^2\right)+\frac34a_0^2\right]
 =\frac32n^2\mathsf M^2.
 \label{eq:worldline-kinetic-completed}
\end{equation}
Combining this result with the static lift ($n=0$) gives the eigenvalue,\footnote{Higher-frequency terms are assumed to contribute only $o(1)$ in this limit. The approximate equality uses the numerical static coefficient.}
\begin{equation}
\begin{aligned}
 \lambda_n
 & =n^2 \frac{\left\langle q_0,\,
 \left(
 \widetilde{\mathcal L}^{\omega}_2
 +\widetilde{\mathcal L}_1
  \widetilde{\mathcal L}_0^{-1}
  \widetilde{\mathcal L}_1
 \right)q_0\right\rangle + \text{cap contribution}}{\langle q_0,q_0\rangle} + \lambda_0 +o(1)\\
 & =  \frac{3 n^2}{2 \mathcal{N}_0}- 1.24156 +o(1)\approx 1.24156(n^2-1).
 \label{eq:worldline-frequency-spectrum}
\end{aligned}
\end{equation}
This calculation explains the numerical spectrum of Section~\ref{sec:localized-modes} through the curvature and frequency corrections to the flat-RN translation, with the horizon contribution fixed by matching to a smooth cap.
To obtain the worldline action, we must now convert the radial normalization to the ultralocal normalization at fixed displacement.

\begin{figure}[t]
 \centering
 \includegraphics[width=0.98\textwidth]{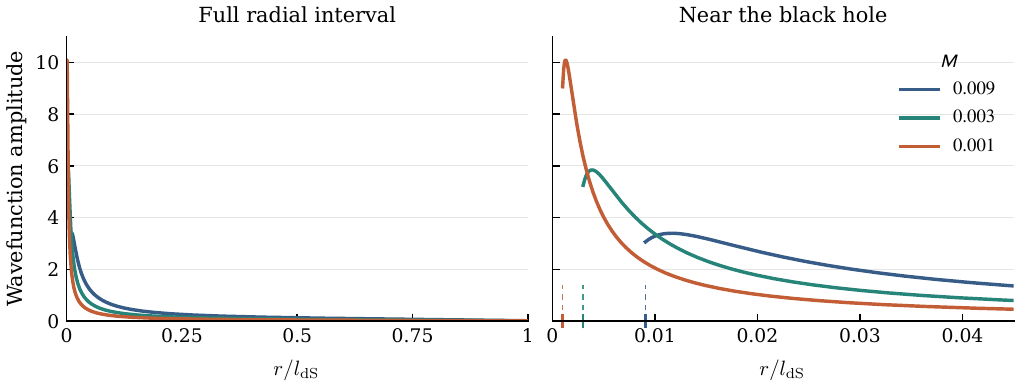}
 \caption{At unit ultralocal norm, the static localized eigenfunction becomes narrower and taller as the black hole shrinks.}
 \label{fig:translation-squeezing}
\end{figure}

\paragraph{The norm at fixed displacement.}
Rescaling an eigenmode leaves its eigenvalue unchanged but changes the displacement it describes.
We impose $P(x),W(x)\to1$, as in Section~\ref{sec:translation-identification}, so that $\boldsymbol Z_n$ is the black hole displacement with units of length.
Let $u_n[\boldsymbol Z_n]$ denote the corresponding metric and Maxwell perturbations.
Their quadratic action is $\lambda_n\|u_n\|_{\rm ultra}^2/(2G_Nl_\text{dS}^2)$, so we need to compute $\|u_n\|_{\rm ultra}^2$ at this fixed displacement.
At fixed $n$ and $x>1$, the leading radial eigenfunction is $q_0(x)$.
Consistently with its localization as $\mathsf M\to0$, we assume that the de Sitter region contributes only a vanishing fraction of the normalization integral.
Restoring $\dd r=l_\text{dS}\mathsf M\dd x$ in \eqref{eq:static-radial-norm} and integrating over time and angles gives
\begin{equation}
 \bigl\|u_n[\boldsymbol Z_n]\bigr\|_{\rm ultra}^2
 =\beta l_\text{dS}\mathsf M\,\frac{2\mathcal N_0}{3}
   |\boldsymbol Z_n|^2\,[1+o(1)].
 \label{eq:worldline-fixed-displacement-norm}
\end{equation}
The time integral gives $\beta$, while $\mathcal N_0\approx1.20816$ is the order-one radial integral in \eqref{eq:app-static-inner-product}.\footnote{For displacement amplitude $\zeta_n$ multiplying a unit-normalized spherical harmonic, the norm is $\beta l_\text{dS}\mathsf M\mathcal N_0|\zeta_n|^2/(2\pi)$. Using $|\boldsymbol Z_n|^2=3|\zeta_n|^2/(4\pi)$ gives \eqref{eq:worldline-fixed-displacement-norm}.}
At fixed displacement, $q_0(x)$ has amplitude independent of $\mathsf M$ and radial width $O(l_\text{dS}\mathsf M)$.
Thus $\|u_n\|_{\rm ultra}^2=O(\mathsf M)$ at fixed $l_\text{dS}$, and its product with the $O(1)$ eigenvalue gives an $O(\mathsf M)$ action, consistent with the worldline theory.
If instead we impose unit ultralocal norm, the weighted amplitudes and the displacement grow as $\mathsf M^{-1/2}$ to compensate for the shrinking width.
Figure~\ref{fig:translation-squeezing} illustrates this behavior for the static eigenfunction.

Substituting \eqref{eq:worldline-frequency-spectrum} and~\eqref{eq:worldline-fixed-displacement-norm} into the quadratic action cancels $\mathcal N_0$.
For a fixed finite set of Fourier harmonics as $\mathsf M\to0$,
\begin{equation*}
 \begin{aligned}
 I^{(2)}[\boldsymbol Z]
 &=\frac{1}{2G_Nl_\text{dS}^2}\sum_n\lambda_n
   \bigl\|u_n[\boldsymbol Z_n]\bigr\|_{\rm ultra}^2\\
 &=\frac{\beta l_\text{dS}\mathsf M}{3G_Nl_\text{dS}^2}
   \sum_n\mathcal N_0\lambda_n|\boldsymbol Z_n|^2
   +o(\mathsf M)\\
 &\simeq\frac{\beta l_\text{dS}\mathsf M}{2G_Nl_\text{dS}^2}
   \sum_n(n^2-1)|\boldsymbol Z_n|^2.
 \end{aligned}
\end{equation*}
{The factor $3/2$ in \eqref{eq:worldline-kinetic-completed} combines with $2/3$ in the norm to give the kinetic coefficient $l_\text{dS}\mathsf M/(2G_N)$.}
Using \eqref{eq:worldline-displacement-frequency} and $l_\text{dS}\omega_n=n+O(\mathsf M)$, we recover the leading worldline action to the accuracy of the static calculation,
\begin{equation}
 I^{(2)}[\boldsymbol Z]
 \simeq\frac{l_\text{dS}\mathsf M}{2G_N}\int_0^\beta\dd\tau
 \left(\dot{\boldsymbol Z}^{\,2}
       -\frac{\boldsymbol Z^{\,2}}{l_\text{dS}^2}\right),
 \qquad \dot{\boldsymbol Z}\equiv\frac{\dd\boldsymbol Z}{\dd\tau}.
 \label{eq:worldline-action-derived}
\end{equation}
Under the matching assumptions above, this is the leading quadratic worldline action for a particle of black hole mass $M=l_\text{dS}\mathsf M/G_N$ in de Sitter.
The potential term agrees with the first-law prediction in Section~\ref{sec:static-translation-lifting}.
At this order, the same mass determines the inertia and gravitational potential energy, as required by the equivalence principle.

\section{Discussion}
\label{sec:discussion}

We studied the coupled graviton--photon spectrum of four-dimensional magnetic lukewarm black holes, and find that the one-loop phase is $-i$ throughout the family.
The sector-by-sector exact diagonalization and count make explicit how the three negative static displacement modes cancel the conformal-mode deficit associated with moving the observer.
The small-mass limit gives us greater analytic control over the signs of these localized eigenvalues:
their eigenvalues remain finite and approach $\lambda_n\simeq1.24(n^2-1)$, so the $n=0$ mode is negative, $n\geq 2$ modes are positive, and $n =1 $ mode has to approach zero from above.
All of these behaviors are as expected for transverse fluctuations of a worldline observer.
Along the way of achieving this, we identified their limiting radial eigenfunctions with the translation zero mode of an extremal asymptotically flat RN black hole and computed its lift by de Sitter curvature.
This calculation requires matching the RN region to the smooth near-horizon cap, whose contribution survives in the leading eigenvalues despite its shrinking radial support.
Combining the static lift, the kinetic term coefficient, and the ultralocal norm at fixed displacement then reproduces the leading worldline action with the physical black hole mass, under the matching assumptions described above.
Together, these results identify the physical picture that the black hole can be viewed as a backreacting observer.

In the small mass limit, one expects factorization of the partition function,  schematically $Z_{\rm bh}\approx Z_{S^4}Z_{\rm Maxwell}Z_{\rm worldline}$. To establish this, we need to compute the ghost spectrum and combine it with the bosonic spectrum studied here to obtain the complete one-loop partition function. This requires consistent determinant regularization and careful treatment of zero modes, light modes, and their integration measures~\cite{Law:2020cpj, Ivo:2025fwe, Ivo:2026ijv}. Another question is how this factorization relates to the bulk and edge decomposition of de Sitter one-loop partition functions~\cite{Anninos:2020hfj,Law:2025ktz}.
It would also be interesting to study an extension to higher dimensions, which would test the cancellation of the dimension-dependent phase for general $D$, beyond $D=4$, given that higher-dimensional lukewarm solutions exist~\cite{Dias:2004px}: a real Euclidean magnetic formulation uses a $(D-2)$-form field strength, dual to the electric Maxwell field.
Finally, in addition to studying small black holes, one can study the limit in which the mass tends to the largest allowed value, which is the Nariai limit \cite{Shi:2025amq,Blacker:2025zca}.
We already know that the phase of the one-loop partition function does not change when we go beyond the lukewarm branch and enter the stable Nariai branch~\cite{Ivo:2025xek,Shi:2026wnk}, then it would be interesting to smoothly track the localized $\ell = 1$ eigenmodes to the Nariai branch and see what they become.

\section*{Acknowledgement}
We thank Shoaib Akhtar and Raghu Mahajan for initial collaboration. We thank Yiming Chen, Victor Ivo, Maciej Kolanowski, Albert Law, Raghu Mahajan, Xiao-Liang Qi, Xiaoyi Shi, Douglas Stanford, Joaquin Turiaci, Chih-Hung Wu and Zimo Sun for helpful discussion. HT especially thanks Douglas Stanford for illuminating conversation at the beginning of this project.
This work is supported in part by the Leinweber Institutes for Theoretical Physics at UC Berkeley and Stanford.
GL is supported by the Department of Energy, Office of Science, Office of High Energy Physics through award DE-SC0025293 and QuantISED award DE-SC0019380.
HT is supported by Shoucheng Zhang Graduate Fellowship.

\appendix
\numberwithin{equation}{section}

\section{Technical details of one-loop computation}
\label{app:active-radial-components}

\paragraph{Active radial amplitudes.}
The cases $\ell=0,1$ and $n=0$ reduce the number of radial amplitudes.
At $\ell=1$, the traceless scalar-derived tensor and the symmetrized derivative of $\mathbb{V}_i^{\ell m}$ vanish, removing $h_{1}^{(\mathsf S)}$ and $h^{(\mathsf V)}$.
At $\ell=0$, all angular derivatives and vector harmonics vanish, leaving only $h_{ab}^{(\mathsf S)}$ and $h_{2}^{(\mathsf S)}$ in the $\mathsf S$ block and $a_{a}^{(\mathsf S)}$ in the $\mathsf V$ block.
For $n\geq1$, these angular reductions apply in each TRS sector: the photon-plus-metric counts are $1+7$ and $3+3$ at $\ell\geq2$, $1+6$ and $3+2$ at $\ell=1$, and $0+4$ and $2+0$ at $\ell=0$, for $\mathsf S$ and $\mathsf V$ respectively.
At $n=0$, $\sin(\omega_0\tau)$ vanishes, so the $\eta=+1$ block retains only components with an even number of $\tau$ indices and the $\eta=-1$ block only those with an odd number.
Combining this rule with the angular reductions gives all the $n=0$ cases in Table~\ref{tab:active-radial-components}.
These are counts of off-shell amplitudes in the gauge-fixed problem, before quotienting gauge transformations, rather than counts of physical polarizations.
There is no further reduction of the harmonic content at $n=1$.

\begin{table}[!htbp]
 \centering
 \small
 \setlength{\tabcolsep}{3pt}
 \renewcommand{\arraystretch}{1.3}
 \begin{tabular}{@{}ccc|c|*{7}{c}|c|*{3}{c}|*{3}{c}|c@{}}
  \hline
  \multicolumn{3}{c|}{Sector}
  & \multicolumn{9}{c|}{$\mathsf S$ block}
  & \multicolumn{7}{c}{$\mathsf V$ block}\\
  \cline{4-19}
  $n$ & $\ell$ & $\eta$ & $a^{(\mathsf V)}$ & $h_{\tau\tau}^{(\mathsf S)}$ & $h_{\tau r}^{(\mathsf S)}$ & $h_{rr}^{(\mathsf S)}$ & $h_{\tau}^{(\mathsf S)}$ & $h_{r}^{(\mathsf S)}$ & $h_{1}^{(\mathsf S)}$ & $h_{2}^{(\mathsf S)}$ & count & $a_{\tau}^{(\mathsf S)}$ & $a_{r}^{(\mathsf S)}$ & $a^{(\mathsf S)}$ & $h_{\tau}^{(\mathsf V)}$ & $h_{r}^{(\mathsf V)}$ & $h^{(\mathsf V)}$ & count\\
  \hline
  $\geq1$ & $\geq2$ & $\pm1$ & $\checkmark$ & $\checkmark$ & $\checkmark$ & $\checkmark$ & $\checkmark$ & $\checkmark$ & $\checkmark$ & $\checkmark$ & $1+7$ & $\checkmark$ & $\checkmark$ & $\checkmark$ & $\checkmark$ & $\checkmark$ & $\checkmark$ & $3+3$\\
  $\geq1$ & $1$ & $\pm1$ & $\checkmark$ & $\checkmark$ & $\checkmark$ & $\checkmark$ & $\checkmark$ & $\checkmark$ &  & $\checkmark$ & $1+6$ & $\checkmark$ & $\checkmark$ & $\checkmark$ & $\checkmark$ & $\checkmark$ &  & $3+2$\\
  $\geq1$ & $0$ & $\pm1$ &  & $\checkmark$ & $\checkmark$ & $\checkmark$ &  &  &  & $\checkmark$ & $0+4$ & $\checkmark$ & $\checkmark$ &  &  &  &  & $2+0$\\
  \hline
  $0$ & $\geq2$ & $+1$ & $\checkmark$ & $\checkmark$ &  & $\checkmark$ &  & $\checkmark$ & $\checkmark$ & $\checkmark$ & $1+5$ &  & $\checkmark$ & $\checkmark$ &  & $\checkmark$ & $\checkmark$ & $2+2$\\
  $0$ & $1$ & $+1$ & $\checkmark$ & $\checkmark$ &  & $\checkmark$ &  & $\checkmark$ &  & $\checkmark$ & $1+4$ &  & $\checkmark$ & $\checkmark$ &  & $\checkmark$ &  & $2+1$\\
  $0$ & $0$ & $+1$ &  & $\checkmark$ &  & $\checkmark$ &  &  &  & $\checkmark$ & $0+3$ &  & $\checkmark$ &  &  &  &  & $1+0$\\
  \hline
  $0$ & $\geq2$ & $-1$ &  &  & $\checkmark$ &  & $\checkmark$ &  &  &  & $0+2$ & $\checkmark$ &  &  & $\checkmark$ &  &  & $1+1$\\
  $0$ & $1$ & $-1$ &  &  & $\checkmark$ &  & $\checkmark$ &  &  &  & $0+2$ & $\checkmark$ &  &  & $\checkmark$ &  &  & $1+1$\\
  $0$ & $0$ & $-1$ &  &  & $\checkmark$ &  &  &  &  &  & $0+1$ & $\checkmark$ &  &  &  &  &  & $1+0$\\
  \hline
 \end{tabular}
 \caption{Active radial amplitudes for each symmetry sector and one angular harmonic.
 A checkmark denotes an active amplitude; a blank entry denotes an absent amplitude.
 The count columns give the number of photon plus metric amplitudes in each block.
 Each $n\geq1$ row applies separately to $\eta=+1$ and $\eta=-1$; the two copies have different time harmonics but identical radial spectra.
 The $n=0$ rows retain only components whose time harmonic is nonzero.}
 \label{tab:active-radial-components}
\end{table}

In particular, the $(n,\ell)=(0,1)$ sector in the $\mathsf S,\eta=+1$ block contains the five amplitudes $a^{(\mathsf V)}$, $h_{\tau\tau}^{(\mathsf S)}$, $h_{rr}^{(\mathsf S)}$, $h_{r}^{(\mathsf S)}$, and $h_{2}^{(\mathsf S)}$ used in Section~\ref{sec:translation-worldline}.
The $n=0$, $\eta=-1$ blocks are nonempty, but contain no metric trace.
The conformal subtraction is therefore needed in the $\mathsf S$ blocks with $n\geq1$ and in the $n=0$, $\mathsf S,\eta=+1$ blocks, but not in the $n=0$, $\eta=-1$ blocks or any $\mathsf V$ block.

\paragraph{Consistency check by modifying gauge-fixing parameters. }
We also checked whether varying the gauge-fixing parameters introduces zero modes or changes the phase at finite mass.
At various different masses, we varied the gauge choice by changing the numerical coefficient for  either term in \eqref{eq:coupled-gauge-fixing}, or replacing the trace coefficient $1/2$ (de Donder gauge) in the metric gauge condition.
For $0\leq n,\ell\leq2$ in both coupled blocks and both time-reflection sectors, no numerical zero modes were detected, and the regularized counts continued to sum to $-3$, giving the same phase $(-\mathrm{i})^{-3}=-\mathrm{i}$.
{These finite-mass checks with other gauge parameters support the phase count in Section~\ref{sec:spectrum-phase}.}

\section{Translation modes and their action}
\label{app:translation-matching}
\addtocontents{toc}{\protect\setcounter{tocdepth}{1}}

This appendix gives the radial calculations underlying the localized-mode eigenvalues and worldline action in Section~\ref{sec:translation-worldline}.
Appendix~\ref{app:expansion-regions} defines the three expansions of the geometry and their overlaps.
Appendix~\ref{app:regions-translation-data} constructs the extremal flat-RN translation, fixes its horizon amplitude $a_0$, and determines its norm at fixed displacement.
In Appendix~\ref{app:static-time-matching} we continue this translation to the smooth black hole horizon and calculate the static eigenvalue correction.
Appendix~\ref{app:nonstatic-matching} gives the additional fields and matching conditions required at nonzero frequency.
Appendix~\ref{app:matched-action} evaluates the frequency-dependent action, retaining the boundary terms.

\subsection{Expansion regions}
\label{app:expansion-regions}

Three expansions describe a small black hole and its displacement in de Sitter.
At distances comparable to the black hole radius, we retain the black hole geometry and treat de Sitter curvature perturbatively.
At larger distances, the geometry approaches de Sitter, which defines the frame in which the displacement is measured.
A third expansion resolves the smooth RN--dS cap near the horizon, where the extremal flat-RN approximation fails.
The relation between matching such expansions and constructing a worldline effective theory is discussed in Ref.~\cite{Kol:2007rx}.

Recall that $\mathsf M=G_NM/l_\text{dS}$, $|Q|=l_\text{dS}\mathsf M$, and the exact lukewarm metric is
\begin{equation}
 \dd s^2=f(r)\dd\tau^2+\frac{\dd r^2}{f(r)}+r^2\dd\Omega_2^2,
 \qquad
 f(r)=\left(1-\frac{l_\text{dS}\mathsf M}{r}\right)^2-\frac{r^2}{l_\text{dS}^2}.
 \label{eq:app-expansion-background}
\end{equation}
Here $r$ is the areal radius, $\tau$ has period $\beta=2\pi l_\text{dS}/\sqrt{1-4\mathsf M}$, and $\dd\Omega_2^2$ is the metric on the unit sphere.
The different approximations follow by taking $\mathsf M\to0$ while holding different radial coordinates fixed.

\paragraph{The RN and de Sitter regions.}
To resolve the geometry on the scale of the black hole radius, set $x=r/(l_\text{dS}\mathsf M)$.
At fixed $x>1$,
\begin{equation}
 f(x)=f_0(x)-\mathsf M^2x^2,
 \qquad f_0(x)=\left(1-\frac1x\right)^2.
 \label{eq:app-flat-rn-expansion}
\end{equation}
Thus the leading geometry is extremal flat-RN, with horizon at $x=1$.
This limit retains the black hole curvature and magnetic field while suppressing the de Sitter correction.
We construct the leading localized eigenfunction on this geometry.

The de Sitter expansion instead holds $r/l_\text{dS}$ fixed.
In these variables, the metric function becomes
\begin{equation}
 f(r)=1-\frac{r^2}{l_\text{dS}^2}-\frac{2l_\text{dS}\mathsf M}{r}
       +\frac{l_\text{dS}^2\mathsf M^2}{r^2}
       \longrightarrow 1-\frac{r^2}{l_\text{dS}^2}.
 \label{eq:app-de-sitter-expansion}
\end{equation}
The leading geometry is de Sitter, and the black hole field is a small correction.
The two approximations overlap when
\begin{equation}
 \mathsf M\ll r/l_\text{dS}\ll1,
 \qquad\hbox{equivalently}\qquad 1\ll x\ll\mathsf M^{-1}.
 \label{eq:app-outer-overlap}
\end{equation}
Both geometries are approximately flat there.
This overlap identifies a translation in the asymptotically flat RN frame with a displacement of the black hole in de Sitter.

\paragraph{Near the black hole horizon.}
The small-$\mathsf M$ expansion in the RN region is not uniform near $x=1$.
There $f_0\simeq(x-1)^2$, so the curvature term in \eqref{eq:app-flat-rn-expansion} is comparable to $f_0$ when $x-1=O(\mathsf M)$.
Both terms are therefore required to locate the physical horizon and impose smoothness.
With $x_h=r_h/(l_\text{dS}\mathsf M)$, introduce a coordinate $X$ measured from the physical horizon:
\begin{equation}
 \begin{gathered}
 X=\frac{l_\text{dS}(r-r_h)}{r_h^2}
   =\frac{x-x_h}{\mathsf M x_h^2},\\
 x_h=1+\mathsf M+2\mathsf M^2+O(\mathsf M^3),
 \qquad x=x_h+\mathsf M x_h^2X,\\
 f(x)=\mathsf M^2F(X)+O(\mathsf M^3),
 \qquad F(X)=X(X+2).
 \end{gathered}
 \label{eq:app-thermal-scaling}
\end{equation}
The physical horizon is at $X=0$, where $F$ has a simple zero.
To express its regularity condition, write $X=\cosh\rho-1$ and $\psi=\kappa_h\tau$, with $\kappa_h=\sqrt{1-4\mathsf M}/l_\text{dS}$.
The angular coordinate $\psi$ has period $2\pi$, and at fixed $\rho$ the metric becomes
\begin{equation}
 \frac{\dd s^2}{l_\text{dS}^2}
 =\mathsf M^2\left(\dd\rho^2+\sinh^2\rho\,\dd\psi^2
                    +\dd\Omega_2^2\right)+O(\mathsf M^3).
 \label{eq:app-smooth-end-metric}
\end{equation}
Near $\rho=0$, the first two terms reduce to the flat polar metric $\dd\rho^2+\rho^2\dd\psi^2$.
Smooth perturbations must therefore be regular tensors at this origin.
The extremal flat-RN throat does not contain this origin, so its boundary condition cannot replace the physical smoothness condition.

The cap and flat-RN expansions overlap for
\begin{equation}
 1\ll X\ll\mathsf M^{-1},
 \qquad\hbox{equivalently}\qquad \mathsf M\ll x-1\ll1.
 \label{eq:app-inner-overlap}
\end{equation}
In this interval, the large-$X$ smooth solution and the small-$(x-1)$ flat-RN solution describe the same fields.
We use this overlap to continue the translation to the actual horizon, and the overlap in \eqref{eq:app-outer-overlap} to hold its displacement relative to the de Sitter frame fixed.
The flat-RN calculation determines the radial solution between these overlaps; the full RN--dS boundary conditions require matching to both adjacent regions.

\subsection{The flat-RN translation mode}
\label{app:regions-translation-data}

We construct the static translation on extremal flat-RN and normalize it by its displacement at infinity.
Horizon regularity selects two branches of the coupled radial equations, and the asymptotic displacement fixes their linear combination.
This determines the leading localized eigenfunction and its radial norm.

\paragraph{Radial equations and boundary conditions.}
In this subsection $x=r/(l_\text{dS}\mathsf M)$, $f_0=(1-1/x)^2$, and a prime denotes $\dd/\dd x$.
Choose a real harmonic $\mathbb{Y}$ with $\widehat{\nabla}^2\mathbb{Y}=-2\mathbb{Y}$ and $\int_{S^2}\dd\Omega\,\mathbb{Y}^2=1$, where $\widehat{\nabla}_i$ is the covariant derivative on the unit sphere.
For a displacement amplitude $\zeta$ with units of length, write the vector generating the translation as
\begin{equation}
 \xi^\tau=0,\qquad
 l_\text{dS}\mathsf M\xi^x=\zeta P(x)\mathbb{Y},\qquad
 l_\text{dS}\mathsf M\xi^i=\zeta\frac{W(x)}x\widehat{\nabla}^i\mathbb{Y}.
 \label{eq:app-translation-generator}
\end{equation}
The functions $P$ and $W$ are dimensionless.
The condition $P(x),W(x)\to1$ at infinity makes this vector approach a Cartesian translation.
For example, $\mathbb{Y}=\sqrt{3/(4\pi)}\cos\theta$ gives $Z_z=\sqrt{3/(4\pi)}\,\zeta$; in general $|\boldsymbol Z|^2=3|\zeta|^2/(4\pi)$.
We keep this displacement fixed when comparing the actions.

The fields $h_{\mu\nu}=\mathcal L_\xi\bar g_{\mu\nu}$ and $a_\mu=(\iota_\xi\bar F)_\mu$ already solve the linearized Einstein--Maxwell equations.
The radial functions are fixed by the gauge conditions of Section~\ref{sec:one-loop}.
The magnetic perturbation is divergence-free on the sphere, so the Maxwell gauge condition is automatic.
The residual de Donder equation, $\nabla^2\xi_\mu+\bar R_{\mu\nu}\xi^\nu=0$, gives
\begin{align}
 0={}&x^2f_0P''+2xf_0P'+4f_0W
 -\left(x^2f_0''+2xf_0'+2f_0+2\right)P,\notag\\
 0={}&x^2f_0W''+\left(x^2f_0'+2xf_0\right)W'
 -\left(xf_0'+2f_0\right)W+2P.
 \label{eq:app-radial-equations}
\end{align}
The two second-order equations admit four independent local solutions.
Set $y=1-1/x$, so the horizon is $y=0$ and infinity is $y\to1$, and let $\alpha=(1+\sqrt{17})/2$.
Two solutions have divergent $P$ or $W$ at the horizon and are excluded.
The remaining branches are admitted by the extremal limit of a smooth horizon.
Their near-horizon expansions follow from \eqref{eq:app-radial-equations}:
\begin{equation}
 \begin{pmatrix}P\\W\end{pmatrix}
 =a_0\begin{pmatrix}
 2y^2-2y^3-\frac53y^4+\cdots\\
 1+y+\frac13y^2+\frac59y^3+\cdots
 \end{pmatrix}
 +a_\alpha y^\alpha
 \begin{pmatrix}1+O(y)\\(\sqrt{17}-5)/4+O(y)\end{pmatrix}.
 \label{eq:app-rn-horizon-data}
\end{equation}
The two terms define horizon solutions $(P^{(0)},W^{(0)})$ and $(P^{(\alpha)},W^{(\alpha)})$, with the displayed normalizations.
The constants $a_0$ and $a_\alpha$ are fixed by the asymptotic conditions.
In particular, $a_0=\lim_{x\to1^+}W(x)$ measures the amplitude of the analytic horizon branch.

The local expansions follow directly from the radial equations, but their relative amplitude requires a global solution.
We continue both horizon solutions numerically to large $x$.
The four independent behaviors there are a growing solution with leading term $x^2(-2,1)^{\mathsf T}$, a constant translation with leading term $(1,1)^{\mathsf T}$, and two decaying solutions beginning at $x^{-1}$ and $x^{-3}$.
We decompose each continued solution in this asymptotic basis.
Let $G_0,G_\alpha$ denote the growing coefficients and $T_0,T_\alpha$ the translation coefficients, with subscripts identifying the two horizon solutions.
The growing solution must cancel, while the translation coefficient must be one:
\begin{equation}
 \begin{pmatrix}G_0&G_\alpha\\T_0&T_\alpha\end{pmatrix}
 \begin{pmatrix}a_0\\a_\alpha\end{pmatrix}
 =\begin{pmatrix}0\\1\end{pmatrix}.
 \label{eq:app-rn-connection-system}
\end{equation}
Solving this linear system with the numerically determined coefficients gives
\begin{equation}
 a_0\approx0.444481,\qquad a_\alpha\simeq0.8074.
 \label{eq:app-rn-connection-data}
\end{equation}
The unit translation at infinity thus fixes $a_0$ independently of the RN--dS eigenvalues.
An independent Chebyshev--Lobatto solution of the boundary-value problem reproduces these radial functions.

\paragraph{The norm at fixed displacement.}
We reconstruct the metric and Maxwell perturbations and evaluate their norm at fixed displacement.
Factor out the same length $\zeta$ and define
$h_{\tau\tau}=\zeta H_{\tau\tau}\mathbb{Y}$,
$h_{xx}=\zeta H_{xx}\mathbb{Y}$,
$h_{xi}=\zeta H_x\widehat{\nabla}_i\mathbb{Y}$,
$h_{ij}=\zeta H_\Omega\Omega_{ij}\mathbb{Y}$, and
$a_i=\zeta A^{(\mathsf V)}\varepsilon_i{}^j\widehat{\nabla}_j\mathbb{Y}$.
Here $\Omega_{ij}$ and $\varepsilon_{ij}$ are the metric and volume form of the unit sphere.
Acting with the vector in \eqref{eq:app-translation-generator} gives
\begin{align}
 l_\text{dS}\mathsf M H_{\tau\tau}&=f_0'P,
 &\frac{H_{xx}}{l_\text{dS}\mathsf M}&=\frac{2P'}{f_0}-\frac{f_0'P}{f_0^2},
 &\frac{H_x}{l_\text{dS}\mathsf M}&=\frac P{f_0}+xW'-W,\notag\\
 \frac{H_\Omega}{l_\text{dS}\mathsf M}&=2x(P-W),
 &A^{(\mathsf V)}&=-\frac{Q}{l_\text{dS}\mathsf M}\frac W{x}.
 \label{eq:app-five-fields}
\end{align}
We take $Q=l_\text{dS}\mathsf M$ below; reversing the charge reverses the Maxwell perturbation and leaves the norm and eigenvalues unchanged.

Recall the positive norm of Section~\ref{sec:one-loop}, \eqref{eq:coupled-ultralocal-inner-product}:
\begin{equation}
 \|u\|_{\rm ultra}^2
 =\frac1{32\pi}\int\dd^4x\,\sqrt{\bar g}
 \left(h_{\mu\nu}^*h^{\mu\nu}+8a_\mu^*a^\mu\right),
 \qquad u=(h_{\mu\nu},a_\mu).
 \label{eq:app-ultralocal-norm}
\end{equation}
The tensor contractions and angular integral assign different weights to the five radial amplitudes.
Absorbing these weights into $q_0$ gives a radial norm that is a sum of squares:
\begin{equation}
 q_0(x)=\frac14
 \begin{pmatrix}
  4A^{(\mathsf V)}\\[2pt]
  \dfrac{2\sqrt{f_0}\,H_x}{l_\text{dS}\mathsf M}\\[4pt]
  \dfrac1{\sqrt2}\left(
   \dfrac{l_\text{dS}\mathsf M xH_{\tau\tau}}{f_0}
   -\dfrac{xf_0H_{xx}}{l_\text{dS}\mathsf M}\right)\\[6pt]
  \dfrac12\left(
   \dfrac{l_\text{dS}\mathsf M xH_{\tau\tau}}{f_0}
   +\dfrac{xf_0H_{xx}}{l_\text{dS}\mathsf M}\right)
   -\dfrac{H_\Omega}{l_\text{dS}\mathsf M x}\\[6pt]
  \dfrac12\left(
   \dfrac{l_\text{dS}\mathsf M xH_{\tau\tau}}{f_0}
   +\dfrac{xf_0H_{xx}}{l_\text{dS}\mathsf M}\right)
   +\dfrac{H_\Omega}{l_\text{dS}\mathsf M x}
 \end{pmatrix}.
 \label{eq:app-flat-norm-components}
\end{equation}
The time integral gives $\beta$, while $\int\dd\Omega\,\mathbb{Y}^2=1$.
With these weights and $\dd r=l_\text{dS}\mathsf M\dd x$, the norm at fixed displacement becomes
\begin{equation}
 \|u[\zeta]\|_{\rm ultra}^2
 =\frac{\beta|\zeta|^2}{2\pi}
   \int_{l_\text{dS}\mathsf M}^{\infty}\dd r\,
       q_0^{\mathsf T}\!\left(\frac r{l_\text{dS}\mathsf M}\right)
       q_0\!\left(\frac r{l_\text{dS}\mathsf M}\right)
 =\frac{\beta l_\text{dS}\mathsf M|\zeta|^2}{2\pi}\,\mathcal N_0,
 \label{eq:app-fixed-displacement-norm}
\end{equation}
where the dimensionless radial inner product and norm are
\begin{equation}
 \langle u,v\rangle=\int_1^\infty\dd x\,u^{\mathsf T}v,
 \qquad
 \mathcal N_0=\langle q_0,q_0\rangle\approx1.20816.
 \label{eq:app-static-inner-product}
\end{equation}
The factor $l_\text{dS}\mathsf M$ comes entirely from the radial width; neither the displacement nor the time integral is included in $\mathcal N_0$.
The same time factor applies to a complex harmonic $e^{\mathrm{i}\omega_n\tau}$, where $\omega_n=2\pi n/\beta$.
The leading radial eigenfunction at fixed $n$ is $q_0$; substituting $|\zeta_n|^2=4\pi|\boldsymbol Z_n|^2/3$ therefore gives \eqref{eq:worldline-fixed-displacement-norm}.
The radial solution thus fixes both $a_0$ and $\mathcal N_0$ at prescribed displacement.

\subsection{Static matching and the eigenvalue lift}
\label{app:static-time-matching}

Equation~\eqref{eq:static-completed-projection} expresses the static eigenvalue in terms of the curvature correction and a horizon contribution:
\begin{equation*}
 \lambda_0
 =\frac{\langle q_0,\mathcal L_2q_0\rangle+\mathcal J_2}{\mathcal N_0}+o(1)\simeq -1.24.
\end{equation*}
We first evaluate the curvature matrix element $\langle q_0,\mathcal L_2q_0\rangle$ in the flat-RN translation mode.
Writing $\mathcal L_2=A_2\partial_x^2+B_2\partial_x+C_2$, the coefficient matrices are
\begin{equation}
 A_2=x^2
 \operatorname{diag}(1,1,1,1,-1),
 \qquad
 B_2=x
 \begin{pmatrix}
 2&\frac1{x-1}&0&0&0\\
 -\frac1{x-1}&2&0&0&0\\
 0&0&2&0&0\\
 0&0&0&2&0\\
 0&0&0&0&-2
 \end{pmatrix}.
 \label{eq:app-A2-B2}
\end{equation}
\begin{equation}
 C_2=
 \begin{pmatrix}
 0&-\frac1{(x-1)^2}&0&0&0\\
 0&-\frac{5x^2-8x+2}{(x-1)^2}&-\frac{\sqrt2x}{x-1}&\frac{2x}{x-1}&0\\
 0&-\frac{\sqrt2x}{x-1}&-\frac{2(x^2+2x-5)}{(x-1)^2}&2\sqrt2&0\\
 0&\frac{2x}{x-1}&2\sqrt2&-4&0\\
 0&0&0&0&8
 \end{pmatrix}.
 \label{eq:app-C2}
\end{equation}
A direct calculation gives
\begin{equation}
 \langle q_0,\mathcal L_2q_0\rangle\approx-1.69756.
 \label{eq:app-fixed-domain-result}
\end{equation}
The curvature contribution alone gives $\lambda_0\simeq-1.40509$, whereas the numerical RN--dS eigenvalue approaches approximately $-1.24$. The difference is accounted for by the horizon contribution $\mathcal{J}_2$.

To calculate $\mathcal J_2$, we solve for the perturbation in the cap and match it to the prescribed RN translation.
The RN action already includes the extremal throat, whose contribution must be replaced by the smooth-cap action.
We then include the boundary term from the integration by parts used to define $\mathcal L_2$.
Together, these contributions give $\mathcal{J}_2 = a_0^2$ and recover $\lambda_0\simeq -1.24$, as we now show.

\paragraph{The smooth cap solution and its matching.}
The cap region has coordinate width $\Delta r=O(l_\text{dS}\mathsf M^2)$ and lapse $f=O(\mathsf M^2)$.
Radial derivative terms in the Lichnerowicz operator scale as $l_\text{dS}^2f\partial_r^2=O(\mathsf M^{-2})$ in this region, so its small coordinate width does not suppress its contribution to the eigenvalue.
As discussed around \eqref{eq:static-cap-power-counting}, the cap equations must therefore be resolved explicitly.
The coordinate $X$ of Appendix~\ref{app:expansion-regions} keeps the physical horizon at $X=0$:
\begin{equation*}
 y=1-\frac1x,\qquad
 x=x_h+\mathsf M x_h^2X,\qquad
 f=\mathsf M^2F+O(\mathsf M^3),\qquad F=X(X+2).
\end{equation*}
Here $x_h=r_h/(l_\text{dS}\mathsf M)=1+\mathsf M+O(\mathsf M^2)$.
The radial and angular functions $P,W$ describe the displacement vector in \eqref{eq:app-translation-generator}.
The RN solution supplies the amplitude $a_0=\lim_{x\to1^+}W(x)$, with
$P\simeq2a_0y^2$ and $W\simeq a_0(1+y)$ near its extremal horizon.
To construct its leading continuation into the cap, where $y=O(\mathsf M)$, we write
$P=\mathsf M^2p(X)+\cdots$ and $W=a_0+\mathsf M w(X)+\cdots$.
These expansions fix the leading matching data; we retain the undetermined field corrections separately when evaluating the action.
Substituting these expansions into the static equations with the physical lapse $f$ gives
\begin{equation}
 Fp''-4p+4a_0F=0,\qquad (Fw')'-2a_0(X+1)=0,
 \label{eq:app-static-smooth-equations}
\end{equation}
where primes denote $\dd/\dd X$ in the cap calculation.

Integrating the second equation twice gives
\begin{equation*}
 Fw'=a_0F+c,\qquad
 w=a_0X+c_0+\frac c2\log\frac{X}{X+2},
\end{equation*}
where $c$ and $c_0$ are integration constants.
The logarithm is singular at $X=0$, so smoothness sets $c=0$.
The first equation has the smooth particular solution $p=2a_0F$.\footnote{The smooth homogeneous solution matches the $y^\alpha$ branch of the RN translation, where $\alpha=(1+\sqrt{17})/2>2$. Its contribution to $P$ begins at order $\mathsf M^\alpha$ at fixed $X$, beyond the order retained here.}
The constant $c_0$ is fixed by matching to the prescribed RN translation.

We now use the overlap to fix $c_0$.
Choose a sphere at $x_m=x_h+\mathsf M x_h^2X_m$ in the overlap $1\ll X_m\ll\mathsf M^{-1}$.
The two radial coordinates at this common sphere obey
\begin{equation}
 y(x_m)=\mathsf M(X_m+1)+O(\mathsf M^2).
 \label{eq:app-common-sphere}
\end{equation}
We first expand in $\mathsf M$ at fixed $X_m$, then take $X_m$ large.
In particular, the two expressions for $W$ are
\begin{equation*}
 \begin{aligned}
 W_{\rm smooth}&=a_0+\mathsf M(a_0X_m+c_0)+\cdots,\\
 W_{\text{flat-RN}}&=a_0+\mathsf M a_0(X_m+1)+\cdots.
 \end{aligned}
\end{equation*}
Matching the constant terms at order $\mathsf M$ fixes $c_0=a_0$.
Thus smoothness and matching together give
\begin{equation}
 P=2a_0\mathsf M^2X(X+2)+o(\mathsf M^2),\qquad
 W=a_0\bigl[1+\mathsf M(X+1)\bigr]+O(\mathsf M^2).
 \label{eq:app-physical-layer-solution}
\end{equation}
For large $X$, the leading terms agree with the flat-RN horizon expansion.
Thus the same amplitude $a_0$, fixed by the displacement at large radius, determines the leading cap solution.

\paragraph{The difference between the cap and throat actions.}
We evaluate the cap action first and then subtract the extremal-throat contribution already included in the RN calculation.
The displacement and physical fields at the matching sphere are identical in the two calculations.
The cap extends from $X=0$ to $X_m$, while the extremal throat extends from $y=0$ to the same sphere at $y_m\equiv y(x_m)$.

We obtain a radial action integral by substituting the static $\ell=1$ perturbations into the gauge-fixed quadratic action in \eqref{eq:full-quadratic-action}, integrating over time and angles, and introducing the weighted fields $q$ of \eqref{eq:app-flat-norm-components}.
After removing the common overall factors discussed in Section~\ref{sec:translation-worldline-matching}, we denote the resulting radial action by $\mathcal I[q]$.
Its density is a sum of terms of the form $[q]_k\partial_x^2[q]_l$, $[q]_k\partial_x[q]_l$, and $[q]_k[q]_l$, where $k,l=1,\ldots,5$ label the radial components.\footnote{For the pure Maxwell kinetic and Lorenz gauge-fixing terms we use $2a_\mu(-\nabla^2a^\mu+\bar R^\mu{}_{\nu}a^\nu)$, which differs from the form in Section~\ref{sec:background-one-loop} by a total derivative. The Einstein--Hilbert and metric--Maxwell mixing terms are kept as displayed.
}

In the operator form used in Section~\ref{sec:static-translation-lifting}, the bulk term is $\langle q,(\mathcal L_0+\mathsf M^2\mathcal L_2+\cdots)q\rangle$, using the inner product in \eqref{eq:static-radial-norm}.
Thus an order-$\mathsf M^2$ change of $\mathcal I$ contributes at order one to the eigenvalue in \eqref{eq:static-curvature-projection}.

We first evaluate the action of the leading matched fields, then show that the undetermined corrections do not affect the result.
We reconstruct the metric and Maxwell perturbations from the displacement in \eqref{eq:app-physical-layer-solution}, using \eqref{eq:app-five-fields} with the cap lapse $f$.
Substituting these fields into this radial density, including the Jacobian $\dd x/\dd X$, gives\footnote{Using the field reconstruction and weights in \eqref{eq:app-five-fields} and~\eqref{eq:app-flat-norm-components}, with the physical lapse $f$, gives $[q]_1=-W/x\simeq-a_0$ and $[q]_4+[q]_5=(x/2)\partial_xP\simeq\mathsf M p'/2$ for the known cap fields. Including $\dd x/\dd X=\mathsf M x_h^2$, substitution into the radial quadratic form cancels its order-$\mathsf M^0$ and order-$\mathsf M^1$ terms and leaves
\begin{equation*}
 \frac{\dd\mathcal I}{\dd X}
 =-\mathsf M^2a_0^2F'
 -\frac{\mathsf M a_0}{2}\frac{\dd}{\dd X}
 \left[F\frac{\dd}{\dd X}\bigl([q]_4+[q]_5\bigr)\right]
 +o(\mathsf M^2).
\end{equation*}
}
\begin{equation}
 \begin{aligned}
 \left.\frac{\dd\mathcal I}{\dd X}\right|_{\text{cap}}
 &=\mathsf M^2\frac{\dd}{\dd X}
   \left[-a_0^2F-\frac{a_0}{4}Fp''\right]+o(\mathsf M^2)\\
 &=-4a_0^2\mathsf M^2(X+1)+o(\mathsf M^2).
 \end{aligned}
\end{equation}
We used $p=2a_0F$, a particular solution of \eqref{eq:app-static-smooth-equations}.
For the extremal throat, the horizon series in \eqref{eq:app-rn-horizon-data} gives
\begin{equation}
 \left.\frac{\dd\mathcal I}{\dd y}\right|_{\mathcal L_0,q_0}
 =-4a_0^2y-\alpha a_0a_\alpha y^{\alpha-1}+O(y^2).
 \label{eq:app-j2-endpoint-action}
\end{equation}
Here $a_\alpha$ multiplies the $y^\alpha$ solution in the RN horizon expansion, with $\alpha=(1+\sqrt{17})/2>2$.
Although $q_0$ is a zero mode, this density retains a total derivative, whose integral over part of the RN interval need not vanish.

Integrating to the common matching sphere gives the cap and extremal-throat contributions.\footnote{The fractional term contributes $O(\mathsf M^{\alpha-2})$ to the coefficient of $\mathsf M^2$ and the analytic remainder contributes $O(\mathsf M)$. Both vanish in the ordered limit.}
The two contributions are
\begin{equation}
 \begin{aligned}
 \text{cap:}\qquad
 \int_0^{X_m}\dd X\,[-4a_0^2\mathsf M^2(X+1)]
 &=-2a_0^2\mathsf M^2\bigl[(X_m+1)^2-1\bigr],\\
 \text{extremal throat:}\qquad
 \int_0^{y_m}\dd y\,[-4a_0^2y]
 &=-2a_0^2y_m^2.
 \end{aligned}
 \label{eq:app-physical-layer-overlap}
\end{equation}
Using $y_m/\mathsf M=X_m+1+o(1)$ from \eqref{eq:app-common-sphere} and taking $\mathsf M\to0$ first at fixed $X_m$, their difference is $2a_0^2\mathsf M^2+o(\mathsf M^2)$.
The terms that grow with $X_m$ cancel, leaving a result independent of where the matching sphere is placed in the overlap.
Although the interval shrinks, its $O(\mathsf M^2)$ radial action produces an order-one eigenvalue shift through the factor $\mathsf M^{-2}$ in $\mathcal O$.

\paragraph{The horizon term in the eigenvalue.}
The coefficient $2a_0^2$ is the difference between the two regional integrals of the density just defined.
The projection in \eqref{eq:static-curvature-projection} instead uses the bulk matrix element $\langle q_0,\mathcal L_2q_0\rangle$.
We must include the surface term relating these two expressions for the quadratic form.

Let $\mathcal I_2[q]$ be the integral of the coefficient of $\mathsf M^2$ in the radial density of $\mathcal I$, expanded at fixed $x$ and radial fields $q$ over the limiting RN interval $1<x<\infty$.
To obtain the mixed first-derivative terms of $\mathcal L_2$, we move a radial derivative between the first two field components by integration by parts.
Second derivatives remain in both forms; keeping the resulting surface term gives
\begin{align}
 \mathcal I_2[q]-\langle q,\mathcal L_2q\rangle
 &=\left[-\frac{x}{x-1}[q]_2[q]_1\right]_{1}^{\infty},
 \label{eq:app-raw-hermitian-identity}\\
 \mathcal I_2[q_0]-\langle q_0,\mathcal L_2q_0\rangle
 &=-a_0^2.
 \label{eq:app-j2-integration-by-parts}
\end{align}
The first identity can be checked directly from the coefficients in \eqref{eq:app-A2-B2}--\eqref{eq:app-C2}.\footnote{Integrating one half of the mixed term by parts gives $-2x[q]_2[q]_1'/(x-1)=\allowbreak x([q]_1[q]_2'-[q]_2[q]_1')/(x-1)-[q]_1[q]_2/(x-1)^2+\allowbreak\partial_x[-x[q]_1[q]_2/(x-1)]$, where primes denote $\dd/\dd x$.}
For the translation, $[q_0]_1\to-a_0$ and $[q_0]_2=a_0(x-1)+o(x-1)$ at the extremal horizon.
The bracket therefore tends to $a_0^2$ at its lower end and vanishes at infinity, giving the second line.
Adding the cap-minus-throat coefficient $2a_0^2$ to $\mathcal I_2[q_0]$ and comparing with \eqref{eq:static-curvature-projection} gives
\begin{equation}
 \begin{aligned}
 \mathcal J_2&=2a_0^2-a_0^2=a_0^2,\\
 \lambda_0
 &=\frac{\langle q_0,\mathcal L_2q_0\rangle+a_0^2}{\langle q_0,q_0\rangle}+o(1)
 \simeq-1.24156+o(1).
 \end{aligned}
 \label{eq:app-static-coefficient}
\end{equation}
The eigenvalue agrees with the small-mass RN--dS spectrum under the outer-region assumptions stated in Section~\ref{sec:static-translation-lifting}.

\paragraph{Why the field corrections do not change the result.}
We now check the effect of the undetermined field corrections on this result.
The weighted fields $q$ have radial inner product $\int\dd x\,q^{\mathsf T}q$.
These rescalings depend on the background lapse, so the cap fields use the physical lapse $f$.
At the common sphere we match the physical tensor components and their radial momenta, defined by the coefficients of field variations in the boundary term of the action.
The sphere merely separates two approximations to a smooth solution and carries no physical source.

Field corrections could contribute at the order retained here.
In the weighted five-field order of \eqref{eq:app-flat-norm-components}, write
\begin{equation}
 [q]_1=-a_0+\mathsf M A_1+\cdots,\qquad
 [q]_4+[q]_5=\frac{\mathsf M}{2}p'+\cdots,
 \qquad \frac12p'=2a_0(X+1).
 \label{eq:app-static-fields-for-action}
\end{equation}
Here $A_1(X)$ is the first Maxwell correction, while the metric combination is already fixed by the smooth cap solution.
We keep $A_1$ unspecified; smoothness requires $A_1'$ to be bounded at $X=0$.

Expand the same radial density in cap coordinates, including the Jacobian $\dd x/\dd X$.
Its coefficients at orders $\mathsf M^0$ and $\mathsf M^1$ vanish, and
\begin{equation}
 \begin{aligned}
 \frac{1}{\mathsf M^2}\frac{\dd\mathcal I}{\dd X}
 &=\frac{\dd}{\dd X}\left[-a_0^2F-\frac{a_0}{4}Fp''+a_0FA_1'\right] + o(1) \\
 &=-4a_0^2(X+1)+\frac{\dd}{\dd X}(a_0FA_1') + o(1).
 \end{aligned}
 \label{eq:app-static-response-current}
\end{equation}
We used $p=2a_0F$, so the two known terms in the first line combine to $-2a_0^2F$.
All other first-order and second-order field corrections cancel from this coefficient.
The dependence on $A_1$ is a total derivative.
Its lower endpoint vanishes because $F(0)=0$ and $A_1'$ is bounded.
At the common sphere the terms involving $A_1$ from the cap and the adjoining RN region are
\begin{equation}
 \begin{aligned}
 \left.\delta\mathcal I_{\rm smooth}\right|_{X_m}
 &=\mathsf M^2a_0FA_1'\big|_{X_m}+o(\mathsf M^2),\\
 \left.\delta\mathcal I_{\rm RN\ outside}\right|_{X_m}
 &=-\mathsf M^2a_0FA_1'\big|_{X_m}+o(\mathsf M^2).
 \end{aligned}
 \label{eq:app-static-interface-cancellation}
\end{equation}
The signs are opposite because the matching sphere is the upper endpoint of the cap interval and the lower endpoint of the adjoining RN interval.
Since the physical fields and momenta match, the two terms cancel.
This cancellation is between the cap and the RN region outside the sphere; the extremal throat inside the sphere is the reference action to be subtracted.
We therefore need neither the full function $A_1(X)$ nor a decay assumption on $A_1'$ in the overlap.
The undetermined correction thus leaves both the cap-minus-throat coefficient and the eigenvalue unchanged.
Consequently, the correction $q_2$ in \eqref{eq:static-response-equation} need not be determined throughout the geometry.

\subsection{Nonzero-frequency matching and the quadratic action}
\label{app:nonstatic-matching}

We now calculate the kinetic term for a time-dependent black hole displacement.
At fixed $n\ne0$, the leading field at $r=O(l_\text{dS}\mathsf M)$ is still the RN translation $q_0$.
Time derivatives source the additional components $h_{\tau x}$ and $h_{\tau i}$, and the expansion about $q_0$ fails at $y=O(|n|\mathsf M)$.
We therefore impose smoothness at the black hole horizon, $X=0$, and match to the RN fields before evaluating their action.
The static contribution was calculated in Appendix~\ref{app:static-time-matching}; here we determine the additional frequency-dependent contribution, including the boundary terms.

\subsubsection{The time-dependent translation and its horizon matching}
\label{app:time-dependent-translation}

The time-dependent calculation requires the matched fields through all orders contributing to the radial action at order $\mathsf M^2$.
We first relate the operators of Section~\ref{sec:translation-worldline-matching} to the five static fields and the two additional metric components, then state the matching results needed below.
We keep $n\ne0$ fixed and use $x=r/(l_\text{dS}\mathsf M)$, $y=1-1/x$, and $X=(x-x_h)/(\mathsf M x_h^2)$ as in Appendix~\ref{app:static-time-matching}, where $x_h$ is the black hole horizon in the $x$ coordinate.
The RN translation $q_0$ has $P(x),W(x)\to1$ as $x\to\infty$ and horizon amplitude $a_0=\lim_{x\to1^+}W(x)$.

\paragraph{The fields and operators.}
The five-component vector $q$ has the radial weights and component order of \eqref{eq:app-flat-norm-components}.
The two-component vector $v$ contains the weighted amplitudes of $h_{\tau x}$ and $h_{\tau i}$.\footnote{After removing the common displacement and harmonic factors as in Appendix~\ref{app:regions-translation-data}, their weights are $x/(2\sqrt2)$ and $1/(2\sqrt f)$, respectively. In the real harmonic convention of Section~\ref{sec:background-one-loop}, $q$ multiplies $\cos(|\omega_n|\tau)$ and $v$ multiplies $\sin(|\omega_n|\tau)$.}
For $n>0$, the corresponding complex Fourier coefficients are $(q,-\mathrm{i}v)^{\mathsf T}$, so the seven-field equation is
\begin{equation}
 \mathsf M^2\mathcal O\binom{q}{-\mathrm{i}v}
 =\mathsf M^2\lambda_n\binom{q}{-\mathrm{i}v}.
 \label{eq:app-nonstatic-master-equation}
\end{equation}
In this basis, the operators in Section~\ref{sec:translation-worldline-matching} have the blocks
\begin{equation}
 \begin{gathered}
 \widetilde{\mathcal L}_0=
 \begin{pmatrix}\mathcal L_0&0\\0&\mathcal H_0\end{pmatrix},
 \qquad
 \widetilde{\mathcal L}_1=
 \begin{pmatrix}0&\mathcal V\\-\mathcal V^{\mathsf T}&0\end{pmatrix},
 \qquad (\widetilde{\mathcal L}_2)_{qq}=\mathcal L_2,\\
 (\widetilde{\mathcal L}_2^\omega)_{qq}
 =\mathcal D=\frac1{y^2}\operatorname{diag}(1,1,1,1,-1).
 \end{gathered}
 \label{eq:newappB-singular-blocks}
\end{equation}
Thus $\mathcal L_0$ and $\mathcal L_2$ are precisely the static five-field operators used in Appendix~\ref{app:static-time-matching}; $\mathcal H_0$ acts on $v$, and $\mathcal V$ couples $v$ to $q$.
The two new operators are
\begin{align}
 \mathcal H_0={}&-\frac{\dd}{\dd x}
 \left[\frac{(x-1)^2}{x^2}\frac{\dd}{\dd x}\right]
 +\begin{pmatrix}
 \dfrac{2(x-1+2x^2)}{x^4}&-\dfrac{2\sqrt2(x-1)}{x^3}\\[2ex]
 -\dfrac{2\sqrt2(x-1)}{x^3}&\dfrac{4+2x(x-1)}{x^4}
 \end{pmatrix},
 \label{eq:newappB-odd-positivity}\\
 \mathcal V={}&\frac{2}{x(x-1)}
 \begin{pmatrix}0&0&-2&0&0\\-1&1&0&0&0\end{pmatrix}^{\mathsf T}.
 \label{eq:app-nonstatic-frequency-coupling}
\end{align}
At fixed $x>1$, the first frequency corrections satisfy
\begin{equation}
 \begin{gathered}
 q=q_0+|n|\mathsf M q_1+\cdots,\qquad
 v=|n|\mathsf M v_1+\cdots,\\
 \mathcal L_0q_1=0,\qquad
 \mathcal H_0v_1=-\mathcal V^{\mathsf T}q_0.
 \end{gathered}
 \label{eq:app-finite-frequency-response}
\end{equation}
This also identifies the second-order expression in Section~\ref{sec:translation-worldline-matching}:
\begin{equation}
 \left(\widetilde{\mathcal L}_2^\omega
 +\widetilde{\mathcal L}_1\widetilde{\mathcal L}_0^{-1}
       \widetilde{\mathcal L}_1\right)_{qq}
 =\mathcal D-\mathcal V\mathcal H_0^{-1}\mathcal V^{\mathsf T}.
 \label{eq:app-frequency-block-reduction}
\end{equation}
Here $\mathcal H_0^{-1}\mathcal V^{\mathsf T}q_0=-v_1$, with $v_1$ fixed by the matching below and by $v_1=O(1)$ as $x\to\infty$.
Using $\mathcal H_0v_1=-\mathcal V^{\mathsf T}q_0$ gives the bulk integrand $q_0^{\mathsf T}\mathcal Dq_0+q_0^{\mathsf T}\mathcal Vv_1$ used in Appendix~\ref{app:matched-action}.

\paragraph{Matching conditions.}
The frequency terms grow as $n^2\mathsf M^2/y^2$, so the expansion at fixed $x$ fails at $y=O(|n|\mathsf M)$.
As in Appendix~\ref{app:static-time-matching}, we match to fields smooth at $X=0$, with $f=\mathsf M^2F+O(\mathsf M^3)$ and $F=X(X+2)$.
Two features of the matching enter the action.
First, we compare the solutions on the same sphere, denoted by $x=x_m$ or $X=X_m$ in the two coordinates:
\begin{equation}
 \frac{y(x_m)}{\mathsf M}
 =X_m+1-\mathsf M(X_m^2-1)+O(\mathsf M^2).
 \label{eq:app-nonstatic-common-sphere}
\end{equation}
We expand first in $\mathsf M$ at fixed $X_m$, then take $X_m\to\infty$.
The subleading term in this coordinate relation changes an order-$\mathsf M$ RN integral at order $\mathsf M^2$ and must be retained.
Second, the weights in $q$ depend on the background metric function, so we match the physical tensor components rather than equating $q_{\rm cap}$ and $q_{\rm RN}$ directly.
For example, matching $h_{xi}$ gives
\begin{equation}
 \frac{[q_{\rm cap}]_2}{\sqrt f}
 =\frac{[q_{\rm RN}]_2}{y}.
 \label{eq:app-common-field-map}
\end{equation}
We differentiate these matching relations when evaluating terms with radial derivatives in the action.
The component equations and matching calculation are given in \texttt{AppendixB.nb}.
Below we list the fields and coefficients required for the action.

\paragraph{The matched fields.}
For $X=O(1)$ in the cap and $y/\mathsf M=O(1)$ in RN, write
\begin{equation}
 q_{\rm cap}=q_{{\rm cap},0}+\mathsf M q_{{\rm cap},1}+\cdots,
 \qquad
 q_{\rm RN}=q_{{\rm RN},0}+\mathsf M q_{{\rm RN},1}+\cdots.
 \label{eq:app-nonstatic-cap-expansion}
\end{equation}
These expansions hold $X$ and $y/\mathsf M$ fixed, respectively. Their first corrections differ from $q_1$, whose expansion holds $x>1$ fixed.
The leading Maxwell amplitudes and induced metric components are
\begin{align}
 a=[q_{{\rm cap},0}]_1
 &=-a_0\left(\frac{X}{X+2}\right)^{|n|/2},
 & [q_{{\rm cap},0}]_2=-[v_{{\rm cap},0}]_2
 &=-\frac{|n|a}{2\sqrt F},
 \label{eq:app-finite-smooth-profile}\\
 a_{\rm RN}=[q_{{\rm RN},0}]_1
 &=-a_0e^{-|n|\mathsf M/y},
 & [q_{{\rm RN},0}]_2=-[v_{{\rm RN},0}]_2
 &=-\frac{|n|\mathsf M a_{\rm RN}}{2y}.
 \label{eq:app-rn-frequency-end}
\end{align}
Here $v_{{\rm cap},0}$ and $v_{{\rm RN},0}$ are the corresponding leading induced fields.
The corresponding physical fields are smooth at $X=0$, while the RN fields decay as $y\to0^+$ at fixed nonzero $|n|\mathsf M$.
Both match the RN translation when $1\ll X\ll\mathsf M^{-1}$.

At the next order, the action depends on the following large-$X$ and large-$y/\mathsf M$ terms. The rows label components or component sums, normalized by the corresponding leading Maxwell amplitude:
\begin{equation}
 \begin{array}{c|cc}
 &q_{{\rm cap},1}/a&q_{{\rm RN},1}/a_{\rm RN}\\ \hline
 1&O(\log X)&O\!\left(\log\frac y{\mathsf M}\right)\\[3pt]
 2&-X+O(\log X)&-\dfrac y{\mathsf M}+O\!\left(\log\frac y{\mathsf M}\right)\\[6pt]
 4+5&-2X+c_{\rm cap}+o(1)&-2\dfrac y{\mathsf M}+c_{\rm RN}+o(1)
 \end{array}
 \label{eq:app-common-response-tails}
\end{equation}
The last row denotes the sum of components four and five, and $c_{\rm cap},c_{\rm RN}$ are their constant terms.
Matching at the same sphere gives
\begin{equation}
 c_{\rm RN}-c_{\rm cap}=2.
 \label{eq:app-nonstatic-matched-constants}
\end{equation}
This shift follows from $y(x_m)/\mathsf M=X_m+1+o(1)$; the individual constants cancel from the action.
The terms proportional to $X$ and $y/\mathsf M$ must also be kept: in particular, $\mathsf M[q_{{\rm RN},1}]_2$ contains $-a_{\rm RN}y$, which leaves a finite surface contribution in Appendix~\ref{app:matched-action}.\footnote{The first corrections include the correction to the physical frequency, $l_\text{dS}|\omega_n|=|n|\sqrt{1-4\mathsf M}$. The displayed RN exponential uses the leading value $l_\text{dS}|\omega_n|=|n|$; the correction is included in $q_{{\rm RN},1}$.}

Re-expanding these matched fields at fixed $y$ gives
\begin{equation}
 q_1=\frac{a_0}{2y}(2,1,0,-1,1)^{\mathsf T}+O(\log y),
 \qquad
 [v_1]_2=-\frac{a_0}{2y}-\frac{3a_0}{2}+o(1)
 \quad(y\to0^+).
 \label{eq:app-response-overlap}
\end{equation}
Thus $\mathcal L_0q_1=0$ does not imply $q_1=0$: its boundary data are fixed by the smooth cap.
The constant term in $[v_1]_2$ is also needed for the finite kinetic integral.
We select $v_1=O(1)$ as $x\to\infty$.

We compare actions at the same displacement and with the same field values at $x=x_o$.\footnote{At finite $x_o>x_m$, we impose $q(x_o)=q_0(x_o)$ and $v(x_o)=0$. We first expand through order $\mathsf M^2$ at fixed $x>1$ and $n$, combine the RN contributions before taking $x\to1^+$, and finally take $x_o\to\infty$.}
As in Section~\ref{sec:translation-worldline-matching}, we assume that the matched fields extend to these boundary values without an additional contribution from $r=O(l_\text{dS})$ at this order.

\subsubsection{The kinetic action and the horizon contribution}
\label{app:matched-action}

We now evaluate the coefficient of $n^2$ in the radial action $\mathcal I$ defined in Appendix~\ref{app:static-time-matching}.
The fields $q_0$ and $v_1$ determine the bulk contribution.
As in Appendix~\ref{app:static-time-matching}, we must also compare the cap and RN actions.
At nonzero frequency there is one further difference: the RN fields decay exponentially as $y\to0^+$, so their expansion at fixed $y$ cannot be used all the way to the horizon.

\paragraph{The bulk contribution.}
In the RN region, time derivatives acting on $q_0e^{\mathrm{i}\omega_n\tau}$ give the term $q_0^{\mathsf T}\mathcal Dq_0$, with $\mathcal D$ defined in \eqref{eq:newappB-singular-blocks}.
The two additional fields $v$ also contribute.
Using $\mathcal H_0v_1=-\mathcal V^{\mathsf T}q_0$, the contribution from $v$ is
$v_1^{\mathsf T}\mathcal H_0v_1+2q_0^{\mathsf T}\mathcal Vv_1=q_0^{\mathsf T}\mathcal Vv_1$, with the associated boundary terms retained in the horizon calculation below.
The bulk contribution to the radial action at order $n^2\mathsf M^2$ is
\begin{equation}
 n^2\mathsf M^2\int_1^\infty\dd x\,
 \left(q_0^{\mathsf T}\mathcal Dq_0+q_0^{\mathsf T}\mathcal Vv_1\right).
 \label{eq:app-kinetic-bulk-integral}
\end{equation}
This is the $\left\langle q_0,\,
 \left(
 \widetilde{\mathcal L}^{\omega}_2
 +\widetilde{\mathcal L}_1
  \widetilde{\mathcal L}_0^{-1}
  \widetilde{\mathcal L}_1
 \right)q_0\right\rangle$ coefficient in \eqref{eq:worldline-kinetic-perturbation}.

The two terms in this integral must be combined before taking $y\to0^+$.
Their near-horizon expansions are
\begin{equation}
 \begin{aligned}
 q_0^{\mathsf T}\mathcal Dq_0
 &=\frac{a_0^2}{y^2}+\frac{2a_0^2}{y}
   +O(y^{\alpha-3})+O(1),\\
 q_0^{\mathsf T}\mathcal Vv_1
 &=-\frac{a_0^2}{y^2}-\frac{2a_0^2}{y}
   +O(y^{\alpha-3})+O(1).
 \end{aligned}
 \label{eq:app-kinetic-singularity-cancellation}
\end{equation}
The $y^{-2}$ and $y^{-1}$ terms cancel. The remaining $y^{\alpha-3}$ terms have finite integrals as $y\to0^+$, since $\alpha=(1+\sqrt{17})/2>2$.

The radial equations express the combined integrand in \eqref{eq:app-kinetic-bulk-integral} as a total derivative, allowing an analytic evaluation.
The induced fields obey the following component equations, with primes denoting $\dd/\dd y$:
\begin{align}
 [v_1]_1
 &=-\frac{1}{2\sqrt2(1-y)}\Bigl[
 \frac{P}{y^2}+\frac{2-3y}{y}W-(1-y)W'+2(1-y)^2\bigl([v_1]_2-y\partial_y[v_1]_2\bigr)\Bigr],\nonumber\\
 0
 &=y^2(1-y)^2\partial_y^2[v_1]_2+2y(1-y)^2\partial_y[v_1]_2 -2\bigl(1+(1-y)^2\bigr)[v_1]_2\nonumber\\
 &\quad +(2y-1)W'+\frac{3y-2}{y(1-y)}W-\frac{P}{y^2(1-y)}.
 \label{eq:newappB-RN-first-odd}
\end{align}
Together with the equations for $P,W$, these relations yield a primitive $\mathcal F$ for the combined integrand:\footnote{An explicit expression, with primes denoting $\dd/\dd y$, is
\begin{equation*}
 \begin{aligned}
 8\mathcal F(y)={}&-\frac{2P^2}{y^3(1-y)}-\frac{2W^2}{1-y}
       +\frac{2(1+7y-6y^2)PW}{y^2(1-y)} +4[v_1]_2\bigl[(3-2y)P+(1-2y)W\bigr]\\
 &+P'\left[\frac P{y^2}+\frac{2-3y}{y}W+2(1-y)^2[v_1]_2\right] -(1-y)P'W'-2y(1-y)^2P'\partial_y[v_1]_2\\
 &+2W'\left[\frac{1-2y}{y}P+W-2y(1-y)[v_1]_2\right]+4(1-y)\partial_y[v_1]_2\bigl[(1-2y)P+yW\bigr].
 \end{aligned}
\end{equation*}
Differentiating and using the $P,W,v_1$ equations verifies this identity.}
\begin{equation}
 \frac{\dd\mathcal F}{\dd y}
 =\frac{q_0^{\mathsf T}\mathcal Dq_0+q_0^{\mathsf T}\mathcal Vv_1}{(1-y)^2},
 \qquad \frac{\dd x}{\dd y}=\frac1{(1-y)^2}.
 \label{eq:app-kinetic-primitive-identity}
\end{equation}

At the extremal horizon, $y\to0^+$ ($x\to1^+$), equations~\eqref{eq:app-rn-horizon-data} and~\eqref{eq:app-response-overlap} give $\mathcal F(0)=3a_0^2/4$.
At RN infinity, $y\to1^-$ ($x\to\infty$), we impose $v_1=O(1)$.
A term proportional to $x$ would produce $l_\text{dS}\mathsf M|\omega_n|x=|\omega_n|r$, changing the leading fields at $r=O(l_\text{dS})$ and the prescribed displacement there.
With this condition,
\begin{equation}
 P=1+p_1(1-y)+\cdots,\qquad
 W=1+(p_1+1)(1-y)+\cdots,\qquad
 [v_1]_2=-\frac{p_1+4}{2}+O(1-y),
 \label{eq:app-kinetic-outer-data}
\end{equation}
where $p_1$ is the coefficient of $1/x$ in $P$, since $1-y=1/x$.
Its value cancels from $\mathcal F(1)=3/2$, so the bulk action is
\begin{equation}
 \eqref{eq:app-kinetic-bulk-integral}= n^2\mathsf M^2\bigl[\mathcal F(1)-\mathcal F(0)\bigr]
 =n^2\mathsf M^2\left(\frac32-\frac34a_0^2\right).
 \label{eq:app-kinetic-bulk-result}
\end{equation}

We next compare the cap and RN actions, keeping the integration-by-parts surface term as in Appendix~\ref{app:static-time-matching}.
The matched cap-minus-RN contribution has no term proportional to $n^2\mathsf M^2$.
The remaining horizon contribution comes from the frequency-dependent RN decay, which the bulk expansion did not retain.

\paragraph{The RN horizon contribution.}
Smoothness around the contractible Euclidean time circle forces the Maxwell amplitude $a$ in \eqref{eq:app-finite-smooth-profile} to vanish at $X=0$ when $n\ne0$.
Its RN continuation contains the decay factor $e^{-|n|\mathsf M/y}$.
The expansion
\begin{equation*}
 e^{-|n|\mathsf M/y}
 =1-\frac{|n|\mathsf M}{y}
  +\frac{n^2\mathsf M^2}{2y^2}+\cdots
\end{equation*}
is valid for $y\gg |n|\mathsf M$ and cannot be integrated term by term down to $y=0$.
The inverse powers of $y$ reflect the rapid radial variation near the horizon at nonzero frequency.
We evaluate the RN action using the matched fields without expanding the factor $e^{-|n|\mathsf M/y}$.
From this we subtract the action calculated by expanding the fields at fixed $y$ before integration, as in the bulk calculation above.
The terms proportional to $|n|\mathsf M$ cancel between the region $y=O(\mathsf M)$ and the remaining RN interval.
The difference at the next order is $n^2\mathsf M^2\mathcal J_\omega$.

To obtain it, we only need the difference between the second-order corrections to the five fields in these two RN calculations.
Write this difference as $n^2\mathsf M^2\Delta q_2^\omega$, subtracting the fields used in the bulk expansion from those retaining the unexpanded factor $e^{-|n|\mathsf M/y}$.
The factor $n^2\mathsf M^2$ is therefore excluded from $\Delta q_2^\omega$. The static curvature correction $q_2$ of Section~\ref{sec:static-translation-lifting} cancels in this comparison.
The two second-order equations have the same bulk source, determined by $q_0$ and $v_1$.
Subtracting them gives
\begin{equation}
 \mathcal L_0\Delta q_2^\omega=0.
 \label{eq:app-comparison-response-equations}
\end{equation}
As in Appendix~\ref{app:static-time-matching}, integrating the cross term with $q_0$ by parts moves $\mathcal L_0$ onto $q_0$.
Since $\mathcal L_0q_0=0$, the bulk term vanishes and the difference is a boundary contribution.
Because both $q_0$ and $\Delta q_2^\omega$ solve the homogeneous RN equation, the resulting boundary expression is independent of $x$.
We evaluate it as $y\to0^+$, where the matching calculation fixes the required coefficients.
With the common boundary data of Appendix~\ref{app:time-dependent-translation}, the coefficient in the action difference is
\begin{equation}
 \mathcal J_\omega
 =-\frac{a_0}{2}\lim_{y\to0^+}
 \left\{y^2\frac{\dd}{\dd y}
 \bigl(2[\Delta q_2^\omega]_1-[\Delta q_2^\omega]_4-[\Delta q_2^\omega]_5\bigr)
       +4y[\Delta q_2^\omega]_2\right\}.
 \label{eq:app-frequency-boundary-pairing}
\end{equation}
Here we used $q_0\to-a_0(2,0,0,-1,1)^{\mathsf T}/2$ as $y\to0^+$.

The terms inside the braces contain logarithmic divergences when evaluated separately.
Adding and subtracting $4y^2\partial_y[\Delta q_2^\omega]_2$ makes the cancellation of logarithms explicit.
The product rule then gives
\begin{equation}
 \mathcal J_\omega
 =-\frac{a_0}{2}\lim_{y\to0^+}
 \left\{\begin{aligned}
 &y^2\frac{\dd}{\dd y}
 \bigl(2[\Delta q_2^\omega]_1-4[\Delta q_2^\omega]_2-[\Delta q_2^\omega]_4-[\Delta q_2^\omega]_5\bigr)\\
 &\qquad +4y\frac{\dd}{\dd y}\bigl(y[\Delta q_2^\omega]_2\bigr)
 \end{aligned}\right\}.
 \label{eq:app-frequency-boundary-rearrangement}
\end{equation}
Both terms now have finite limits as $y\to0^+$.
The detailed calculation, recorded in \texttt{AppendixB.nb}, gives
\begin{equation}
 \mathcal J_\omega
 =-\frac{a_0}{2}\left[-\frac{11a_0}{2}
 +4 a_0
 \right]
 =\frac34a_0^2.
 \label{eq:app-frequency-endpoint-result}
\end{equation}

Adding this horizon term to the bulk action gives
\begin{equation}
 \begin{gathered}
 \mathsf M^2 \mathcal{I}\big|_{n^2} = n^2\mathsf M^2\left[
 \left\langle q_0,\,
 \left(
 \widetilde{\mathcal L}^{\omega}_2
 +\widetilde{\mathcal L}_1
  \widetilde{\mathcal L}_0^{-1}
  \widetilde{\mathcal L}_1
 \right)q_0\right\rangle+\mathcal J_\omega\right] =n^2\mathsf M^2\left[
 \left(\frac32-\frac34a_0^2\right)+\frac34a_0^2\right]
 =\frac32n^2\mathsf M^2,\\
 \lambda_n=\lambda_0+\frac{3n^2}{2\mathcal N_0}+o(1) \simeq 1.24 (n^2 -1).
 \end{gathered}
 \label{eq:app-completed-frequency-coefficient}
\end{equation}
Here $\mathcal N_0=\int_1^\infty\dd x\,q_0^{\mathsf T}q_0$ is the radial norm in \eqref{eq:app-static-inner-product}.

For $|n|=1$, the leading static and kinetic terms cancel to the accuracy of the static calculation.
We next determine the sign of the first nonzero correction to the $n=1$ eigenvalue.

\subsubsection{The \texorpdfstring{$n=1$}{n=1} eigenvalue at order \texorpdfstring{$\mathsf M$}{M}}
\label{app:n1-displacement-action}

For $n=1$, the leading static and kinetic terms cancel.
We now calculate the first nonzero correction to the eigenvalue of the displacement introduced in Section~\ref{sec:localized-modes}.
The static and kinetic terms above were obtained by matching the RN region to the cap.
At the next order, we also need to match the RN translation to the de Sitter region, $r=O(l_\text{dS})$.

In empty de Sitter space the $n=1$ displacement is an isometry and produces no metric perturbation.
The black hole breaks this symmetry, giving a perturbation of order $\mathsf M$ at fixed $r/l_\text{dS}$.
Its gauge-fixing action is of order $\mathsf M^2$, whereas $\|u\|_{\rm ultra}^2$ at fixed displacement is of order $\mathsf M$ and is dominated by the RN region.
Their quotient therefore contributes at order $\mathsf M$ to the eigenvalue.
We calculate the de Sitter contribution with its amplitude fixed by matching to the RN fields.

\paragraph{The displacement vector.}
The displacement is the time-dependent continuation of the RN translation in \eqref{eq:app-translation-generator}.
At fixed $x>1$, its leading spatial components are those of that vector multiplied by $e^{\mathrm{i}\psi}$, where $\psi=\omega_1\tau=2\pi\tau/\beta$.
The first corrections include a time component and are fixed by smoothness at $X=0$ and matching to the de Sitter fields at $r=O(l_\text{dS})$.
To calculate the action from $r=O(l_\text{dS})$, set $l_\text{dS}=1$, write $R=r/l_\text{dS}$, and expand at fixed $(R,\psi)$.
The matched vector has the form
\begin{equation}
 \begin{aligned}
 \xi&=\zeta\bigl(K+\mathsf M\xi_1+\cdots\bigr),\\
 K&=e^{\mathrm{i}\psi}\left[
 -\frac{\mathrm{i}R}{\sqrt{1-R^2}}\mathbb{Y}\,\partial_\psi
 +\sqrt{1-R^2}\mathbb{Y}\,\partial_R
 +\frac{\sqrt{1-R^2}}R \widehat{\nabla}^i\mathbb{Y}\,\partial_i
 \right].
 \end{aligned}
 \label{eq:app-n1-displacement-vector}
\end{equation}
Here $\zeta$ is the displacement amplitude of Appendix~\ref{app:regions-translation-data}, $\mathbb{Y}$ is the same $\ell=1$ harmonic with $\int_{S^2}\dd\Omega\,|\mathbb{Y}|^2=1$, and $\widehat{\nabla}_i$ is the covariant derivative on the unit sphere.
The vector $K$ is a Killing vector of empty de Sitter space.
As $R\to0$, its spatial part approaches $e^{\mathrm{i}\psi}(\mathbb{Y}\partial_R+R^{-1}\widehat{\nabla}^i\mathbb{Y}\partial_i)$, which agrees with the unit RN translation in the overlap $\mathsf M\ll R\ll1$.
We choose $\xi_1$ so that the generated fields satisfy the leading equations at fixed $R$, match $q_1,v_1$ in Appendix~\ref{app:time-dependent-translation}, and are smooth at the cosmological horizon, $R=1$.

The generated fields are $h^\xi=\mathcal L_\xi\bar g$ and $a^\xi_\mu=\xi^\nu\bar F_{\nu\mu}$, as in Section~\ref{sec:translation-identification}, where $\bar g$ and $\bar F$ are the background metric and Maxwell field strength. We denote the pair by $u_\xi=(h^\xi,a^\xi)$.
Writing $\bar g=g_0+\mathsf M g_1+\cdots$ in these same coordinates, with $g_0$ the unit de Sitter metric, gives
\begin{equation}
 h^\xi=\zeta\mathsf M h_1+o(\mathsf M),\qquad
 h_1=\mathcal L_Kg_1+\mathcal L_{\xi_1}g_0.
 \label{eq:app-n1-displacement-fields}
\end{equation}
The first nonzero metric perturbation includes both the black hole correction to the background and the correction to the displacement vector; the leading term vanishes because $\mathcal L_Kg_0=0$.

\paragraph{The gauge condition and RN matching.}
For a displacement, \eqref{eq:n1-displacement-positive-action} reduces the action to the gauge-fixing term.
The Maxwell perturbation is divergence-free, as in Appendix~\ref{app:regions-translation-data}, so only the de Donder term contributes.
The de Donder gauge-fixing function is $\zeta\mathsf M C_{1\mu}+o(\mathsf M)$, with
\begin{equation}
 C_{1\mu}=\nabla^\nu\left(h_{1\nu\mu}-\frac12g_{0\nu\mu}h_1\right),
 \qquad h_1=g_0^{\mu\nu}h_{1\mu\nu},
 \label{eq:app-n1-gauge-condition}
\end{equation}
Here derivatives and index operations use $g_0$.
The quantity $C_1$ measures the leading failure of the displacement to satisfy de Donder gauge; its squared norm determines the action.

We can determine $C_1$ without solving for every component of $\xi_1$.
Since the eigenvalue tends to zero and $h^\xi=O(\mathsf M)$ at fixed $0<R<1$, the eigenvalue term is smaller than order $\mathsf M$.
At that order, the linearized Einstein--Maxwell equations without gauge fixing are satisfied by the displacement, leaving the gauge-fixing equation
\begin{equation}
 -\nabla_{(\mu}C_{1\nu)}
 +\frac12g_{0\mu\nu}\nabla_\rho C_1^\rho=0.
 \label{eq:app-n1-residual-equation}
\end{equation}
Taking its trace gives $\nabla_\rho C_1^\rho=0$, so the equation reduces to $\nabla_{(\mu}C_{1\nu)}=0$. Thus $C_1^\mu$ satisfies the Killing equation.
In the chosen $n=1$, $\ell=1$ harmonic, it must therefore have the form $C_{1\mu}=A_KK_\mu$, with $K_\mu=g_{0\mu\nu}K^\nu$ and a constant $A_K$.
The action depends only on $A_K$, which RN matching fixes without determining the full correction $\xi_1$.

The RN translation fixes $A_K$ through its behavior as $x\to\infty$, which we have already used above,
\begin{equation*}
 P=1+\frac{p_1}{x}+\cdots,\qquad
 W=1+\frac{p_1+1}{x}+\cdots.
\end{equation*}
Since $x=R/\mathsf M$, these $1/x$ terms give the order-$\mathsf M/R$ terms of the displacement at fixed $R$.
Integrating $\overline K^\mu C_{1\mu}$ by parts gives a boundary term; the volume integral vanishes because $\nabla_{(\mu}K_{\nu)}=0$.
The boundary term at the smooth cosmological horizon, $R=1$, vanishes, while matching to the RN fields as $R\to0^+$ gives $3p_1$ after removing the common time integral.
The explicit vector in \eqref{eq:app-n1-displacement-vector} has $\int_{S^2}\dd\Omega\,|K|^2=3-R^2$.
Consequently,
\begin{equation}
 A_K\int_0^1\dd R\,R^2(3-R^2)
 =\frac45A_K=3p_1,
 \qquad A_K=\frac{15}{4}p_1.
 \label{eq:app-n1-nonzero-residual}
\end{equation}
The $1/x$ tail of the RN translation thus fixes the nonzero de Donder gauge-fixing function in the de Sitter region.
Its squared norm gives the positive action that remains after the leading static and kinetic terms cancel.

\paragraph{The action and eigenvalue.}
Substituting $C_1=A_KK$ into the gauge-fixing action, \eqref{eq:coupled-gauge-fixing}, gives the contribution from $0<R<1$:
\begin{equation}
 \begin{aligned}
 I_{\rm gf,ext}^{(2)}
 &=\frac{\beta|\zeta|^2\mathsf M^2}{32\pi G_N}
   \left(\frac{15p_1}{4}\right)^2
   \int_0^1\dd R\,R^2(3-R^2)+o(\mathsf M^2),\\
 \frac{2G_N I_{\rm gf,ext}^{(2)}}{\|u_\xi\|_{\rm ultra}^2}
 &=\frac{\mathsf M}{\mathcal N_0}
   \frac18\left(\frac{15p_1}{4}\right)^2\frac45
   +o(\mathsf M)
 =\frac{45p_1^2}{32\mathcal N_0}\mathsf M+o(\mathsf M).
 \end{aligned}
 \label{eq:app-n1-action-coefficient}
\end{equation}
Here we used the norm at fixed displacement from \eqref{eq:app-fixed-displacement-norm},
$\|u_\xi\|_{\rm ultra}^2=\beta|\zeta|^2\mathsf M\mathcal N_0/(2\pi)+o(\mathsf M)$ in units $l_\text{dS}=1$.
The time integral and displacement amplitude cancel in the quotient; the factor $1/8$ follows from the gauge-fixing and norm prefactors.

To identify this quotient with the first nonzero eigenvalue, we assume that both the remaining gauge-fixing action near the black hole, including the smooth cap, and the error in the quadratic action from replacing the eigenmode by the matched displacement are $o(\mathsf M^2)$.
Matching the leading fields supports this approximation but does not prove that these action errors are $o(\mathsf M^2)$.
With this assumption, the static RN data $p_1\approx-1.70322$ and $\mathcal N_0\approx1.20816$ give
\begin{equation}
 \lambda_1
 =\frac{45p_1^2}{32\mathcal N_0}\mathsf M+o(\mathsf M)
 \simeq3.37662\,\mathsf M+o(\mathsf M).
 \label{eq:app-n1-action-approximation}
\end{equation}
The eigenvalue is dimensionless, as throughout the paper, so restoring $l_\text{dS}$ does not change this formula.
Both inputs to its coefficient come from the static RN translation calculation.
The positive sign predicts that the $n=1$ eigenvalue approaches zero from above, in agreement with Section~\ref{sec:localized-modes}.

\addtocontents{toc}{\protect\setcounter{tocdepth}{2}}

\bibliography{references}

\providecommand{\href}[2]{#2}\begingroup\raggedright\begin{thebibliography}{10}

\bibitem{Gibbons:1976ue}
G.~W. Gibbons and S.~W. Hawking, ``{Action Integrals and Partition Functions in
  Quantum Gravity},'' \href{http://dx.doi.org/10.1103/PhysRevD.15.2752}{{\em
  Phys. Rev. D} {\bfseries 15} (1977) 2752--2756}.

\bibitem{Gibbons:1977mu}
G.~W. Gibbons and S.~W. Hawking, ``{Cosmological Event Horizons,
  Thermodynamics, and Particle Creation},''
  \href{http://dx.doi.org/10.1103/PhysRevD.15.2738}{{\em Phys. Rev. D}
  {\bfseries 15} (1977) 2738--2751}.

\bibitem{Polchinski:1988ua}
J.~Polchinski, ``{The phase of the sum over spheres},''
  \href{http://dx.doi.org/10.1016/0370-2693(89)90387-0}{{\em Phys. Lett. B}
  {\bfseries 219} (1989) 251--257}.

\bibitem{Gibbons:1978ac}
G.~W. Gibbons, S.~W. Hawking, and M.~J. Perry, ``{Path Integrals and the
  Indefiniteness of the Gravitational Action},''
  \href{http://dx.doi.org/10.1016/0550-3213(78)90161-X}{{\em Nucl. Phys. B}
  {\bfseries 138} (1978) 141--150}.

\bibitem{Law:2020cpj}
Y.~T.~A. Law, ``{A compendium of sphere path integrals},''
  \href{http://dx.doi.org/10.1007/JHEP12(2021)213}{{\em JHEP} {\bfseries 12}
  (2021) 213}, \href{http://arxiv.org/abs/2012.06345}{{\ttfamily
  arXiv:2012.06345 [hep-th]}}.

\bibitem{Law:2025yec}
Y.~T.~A. Law and V.~Lochab, ``{Gravitons on Nariai edges},''
  \href{http://dx.doi.org/10.1103/23vk-l1fl}{{\em Phys. Rev. D} {\bfseries 113}
  no.~8, (2026) 086006}, \href{http://arxiv.org/abs/2506.02142}{{\ttfamily
  arXiv:2506.02142 [hep-th]}}.

\bibitem{Maldacena:2024spf}
J.~Maldacena, ``{Real observers solving some imaginary problems},''
  \href{http://dx.doi.org/10.1142/s0217751x2647010x}{{\em Int. J. Mod. Phys. A}
  {\bfseries 41} no.~26, (2026) 2647010},
  \href{http://arxiv.org/abs/2412.14014}{{\ttfamily arXiv:2412.14014
  [hep-th]}}.

\bibitem{Chandrasekaran:2022cip}
V.~Chandrasekaran, R.~Longo, G.~Penington, and E.~Witten, ``{An algebra of
  observables for de Sitter space},''
  \href{http://dx.doi.org/10.1007/JHEP02(2023)082}{{\em JHEP} {\bfseries 02}
  (2023) 082}, \href{http://arxiv.org/abs/2206.10780}{{\ttfamily
  arXiv:2206.10780 [hep-th]}}.

\bibitem{Witten:2023xze}
E.~Witten, ``{A background-independent algebra in quantum gravity},''
  \href{http://dx.doi.org/10.1007/JHEP03(2024)077}{{\em JHEP} {\bfseries 03}
  (2024) 077}, \href{http://arxiv.org/abs/2308.03663}{{\ttfamily
  arXiv:2308.03663 [hep-th]}}.

\bibitem{Witten:2026whn}
E.~Witten, ``{An observer in de Sitter space, and rereading Everett},''
  \href{http://dx.doi.org/10.1142/s0217751x26470093}{{\em Int. J. Mod. Phys. A}
  {\bfseries 41} no.~26, (2026) 2647009},
  \href{http://arxiv.org/abs/2610.04804}{{\ttfamily arXiv:2610.04804
  [hep-th]}}.

\bibitem{Abdalla:2025gzn}
A.~I. Abdalla, S.~Antonini, L.~V. Iliesiu, and A.~Levine, ``{The gravitational
  path integral from an observer{\textquoteright}s point of view},''
  \href{http://dx.doi.org/10.1007/JHEP05(2025)059}{{\em JHEP} {\bfseries 05}
  (2025) 059}, \href{http://arxiv.org/abs/2501.02632}{{\ttfamily
  arXiv:2501.02632 [hep-th]}}.

\bibitem{Harlow:2025pvj}
D.~Harlow, M.~Usatyuk, and Y.~Zhao, ``{Quantum mechanics and observers for
  gravity in a closed universe},''
  \href{http://dx.doi.org/10.1007/JHEP02(2026)108}{{\em JHEP} {\bfseries 02}
  (2026) 108}, \href{http://arxiv.org/abs/2501.02359}{{\ttfamily
  arXiv:2501.02359 [hep-th]}}.

\bibitem{Kolchmeyer:2024fly}
D.~K. Kolchmeyer and H.~Liu, ``{Chaos and the Emergence of the Cosmological
  Horizon},'' \href{http://arxiv.org/abs/2411.08090}{{\ttfamily
  arXiv:2411.08090 [hep-th]}}.

\bibitem{Milekhin:2026tbi}
A.~Milekhin, V.~Narovlansky, and J.~Xu, ``{Out-of-time-ordered Correlators in
  de Sitter Revisited},'' \href{http://arxiv.org/abs/2607.13137}{{\ttfamily
  arXiv:2607.13137 [hep-th]}}.

\bibitem{Cui:2026bcd}
W.~Cui and D.~K. Kolchmeyer, ``{A de Sitter Anti-Scrambling Algebra},''
  \href{http://arxiv.org/abs/2607.13665}{{\ttfamily arXiv:2607.13665
  [hep-th]}}.

\bibitem{Chen:2026boh}
Y.~Chen, D.~Stanford, H.~Tang, and Z.~Yang, ``{Negative shocks versus static
  patch holography},'' \href{http://arxiv.org/abs/2607.14042}{{\ttfamily
  arXiv:2607.14042 [hep-th]}}.

\bibitem{Harlow:2026pwe}
D.~Harlow and Y.~Zhao, ``{Anti-scrambling and euclidean folds from observer
  correlators in de Sitter space},''
  \href{http://arxiv.org/abs/2607.14215}{{\ttfamily arXiv:2607.14215
  [hep-th]}}.

\bibitem{Ivo:2025yek}
V.~Ivo, J.~Maldacena, and Z.~Sun, ``{Physical instabilities and the phase of
  the Euclidean path integral},''
  \href{http://dx.doi.org/10.1007/JHEP04(2026)118}{{\em JHEP} {\bfseries 04}
  (2026) 118}, \href{http://arxiv.org/abs/2504.00920}{{\ttfamily
  arXiv:2504.00920 [hep-th]}}.

\bibitem{Shi:2025amq}
X.~Shi and G.~J. Turiaci, ``{The phase of the gravitational path integral},''
  \href{http://dx.doi.org/10.1007/JHEP07(2025)047}{{\em JHEP} {\bfseries 07}
  (2025) 047}, \href{http://arxiv.org/abs/2504.00900}{{\ttfamily
  arXiv:2504.00900 [hep-th]}}.

\bibitem{Ivo:2025xek}
V.~Ivo and Z.~Sun, ``{The phase of charged Nariai solutions},''
  \href{http://arxiv.org/abs/2511.06604}{{\ttfamily arXiv:2511.06604
  [hep-th]}}.

\bibitem{Ivo:2025fwe}
V.~Ivo, ``{One loop aspects of Coleman de Luccia instantons at small
  backreaction},'' \href{http://arxiv.org/abs/2509.18651}{{\ttfamily
  arXiv:2509.18651 [hep-th]}}.

\bibitem{Ivo:2026ijv}
V.~Ivo and H.~Tang, ``{One-loop aspects of de Sitter axion wormholes},''
  \href{http://arxiv.org/abs/2603.02335}{{\ttfamily arXiv:2603.02335
  [hep-th]}}.

\bibitem{Giombi:2026sqa}
S.~Giombi and Z.~Sun, ``{The phase of de Sitter higher spin gravity},''
  \href{http://arxiv.org/abs/2601.15257}{{\ttfamily arXiv:2601.15257
  [hep-th]}}.

\bibitem{Chen:2025jqm}
Y.~Chen, D.~Stanford, H.~Tang, and Z.~Yang, ``{On the phase of the de Sitter
  density of states},'' \href{http://dx.doi.org/10.1007/JHEP05(2026)068}{{\em
  JHEP} {\bfseries 05} (2026) 068},
  \href{http://arxiv.org/abs/2511.01400}{{\ttfamily arXiv:2511.01400
  [hep-th]}}.

\bibitem{Marolf:2022ybi}
D.~Marolf, ``{Gravitational thermodynamics without the conformal factor
  problem: partition functions and Euclidean saddles from Lorentzian path
  integrals},'' \href{http://dx.doi.org/10.1007/JHEP07(2022)108}{{\em JHEP}
  {\bfseries 07} (2022) 108}, \href{http://arxiv.org/abs/2203.07421}{{\ttfamily
  arXiv:2203.07421 [hep-th]}}.

\bibitem{Kolanowski:2026gii}
M.~Kolanowski and D.~Marolf, ``{How to tame your (black hole) saddles: Lessons
  from the Lorentzian Gravitational Path Integral},''
  \href{http://arxiv.org/abs/2603.24681}{{\ttfamily arXiv:2603.24681
  [hep-th]}}.

\bibitem{Marolf:1996gb}
D.~Marolf, ``{Path integrals and instantons in quantum gravity: Minisuperspace
  models},'' \href{http://dx.doi.org/10.1103/PhysRevD.53.6979}{{\em Phys. Rev.
  D} {\bfseries 53} (1996) 6979--6990},
  \href{http://arxiv.org/abs/gr-qc/9602019}{{\ttfamily arXiv:gr-qc/9602019}}.

\bibitem{Casali:2021ewu}
E.~Casali, D.~Marolf, H.~Maxfield, and M.~Rangamani, ``{Baby universes and
  worldline field theories},''
  \href{http://dx.doi.org/10.1088/1361-6382/ac37cd}{{\em Class. Quant. Grav.}
  {\bfseries 39} no.~13, (2022) 134004},
  \href{http://arxiv.org/abs/2101.12221}{{\ttfamily arXiv:2101.12221
  [hep-th]}}.

\bibitem{Mann:1995vb}
R.~B. Mann and S.~F. Ross, ``{Cosmological production of charged black hole
  pairs},'' \href{http://dx.doi.org/10.1103/PhysRevD.52.2254}{{\em Phys. Rev.
  D} {\bfseries 52} (1995) 2254--2265},
  \href{http://arxiv.org/abs/gr-qc/9504015}{{\ttfamily arXiv:gr-qc/9504015}}.

\bibitem{Romans:1991nq}
L.~J. Romans, ``{Supersymmetric, cold and lukewarm black holes in cosmological
  Einstein-Maxwell theory},''
  \href{http://dx.doi.org/10.1016/0550-3213(92)90684-4}{{\em Nucl. Phys. B}
  {\bfseries 383} (1992) 395--415},
  \href{http://arxiv.org/abs/hep-th/9203018}{{\ttfamily arXiv:hep-th/9203018}}.

\bibitem{Booth:1998gf}
I.~S. Booth and R.~B. Mann, ``{Cosmological pair production of charged and
  rotating black holes},''
  \href{http://dx.doi.org/10.1016/S0550-3213(98)00756-1}{{\em Nucl. Phys. B}
  {\bfseries 539} (1999) 267--306},
  \href{http://arxiv.org/abs/gr-qc/9806056}{{\ttfamily arXiv:gr-qc/9806056}}.

\bibitem{Morvan:2022aon}
E.~K. Morvan, J.~P. van~der Schaar, and M.~R. Visser, ``{Action, entropy and
  pair creation rate of charged black holes in de Sitter space},''
  \href{http://dx.doi.org/10.1007/JHEP08(2024)207}{{\em JHEP} {\bfseries 08}
  (2024) 207}, \href{http://arxiv.org/abs/2212.12713}{{\ttfamily
  arXiv:2212.12713 [hep-th]}}.

\bibitem{Castro:2022cuo}
A.~Castro, F.~Mariani, and C.~Toldo, ``{Near-extremal limits of de Sitter black
  holes},'' \href{http://dx.doi.org/10.1007/JHEP07(2023)131}{{\em JHEP}
  {\bfseries 07} (2023) 131}, \href{http://arxiv.org/abs/2212.14356}{{\ttfamily
  arXiv:2212.14356 [hep-th]}}.

\bibitem{Karan:2021teq}
S.~Karan and B.~Panda, ``{Generalized Einstein-Maxwell theory: Seeley-DeWitt
  coefficients and logarithmic corrections to the entropy of extremal and
  nonextremal black holes},''
  \href{http://dx.doi.org/10.1103/PhysRevD.104.046010}{{\em Phys. Rev. D}
  {\bfseries 104} no.~4, (2021) 046010},
  \href{http://arxiv.org/abs/2104.06381}{{\ttfamily arXiv:2104.06381
  [hep-th]}}.

\bibitem{Kodama:2003kk}
H.~Kodama and A.~Ishibashi, ``{Master equations for perturbations of
  generalized static black holes with charge in higher dimensions},''
  \href{http://dx.doi.org/10.1143/PTP.111.29}{{\em Prog. Theor. Phys.}
  {\bfseries 111} (2004) 29--73},
  \href{http://arxiv.org/abs/hep-th/0308128}{{\ttfamily arXiv:hep-th/0308128}}.

\bibitem{Trefethen2000Spectral}
L.~N. Trefethen, \href{http://dx.doi.org/10.1137/1.9780898719598}{{\em Spectral
  Methods in {MATLAB}}}, vol.~10 of {\em Software, Environments, and Tools}.
\newblock SIAM, Philadelphia, PA, 2000.

\bibitem{Kolanowski:2024zrq}
M.~Kolanowski, D.~Marolf, I.~Rakic, M.~Rangamani, and G.~J. Turiaci, ``{Looking
  at extremal black holes from very far away},''
  \href{http://dx.doi.org/10.1007/JHEP04(2025)020}{{\em JHEP} {\bfseries 04}
  (2025) 020}, \href{http://arxiv.org/abs/2409.16248}{{\ttfamily
  arXiv:2409.16248 [hep-th]}}.

\bibitem{Gross:1982cv}
D.~J. Gross, M.~J. Perry, and L.~G. Yaffe, ``{Instability of Flat Space at
  Finite Temperature},'' \href{http://dx.doi.org/10.1103/PhysRevD.25.330}{{\em
  Phys. Rev. D} {\bfseries 25} (1982) 330--355}.

\bibitem{Goldberger:2004jt}
W.~D. Goldberger and I.~Z. Rothstein, ``{An Effective field theory of gravity
  for extended objects},''
  \href{http://dx.doi.org/10.1103/PhysRevD.73.104029}{{\em Phys. Rev. D}
  {\bfseries 73} (2006) 104029},
  \href{http://arxiv.org/abs/hep-th/0409156}{{\ttfamily arXiv:hep-th/0409156}}.

\bibitem{Kol:2007rx}
B.~Kol and M.~Smolkin, ``{Classical Effective Field Theory and Caged Black
  Holes},'' \href{http://dx.doi.org/10.1103/PhysRevD.77.064033}{{\em Phys. Rev.
  D} {\bfseries 77} (2008) 064033},
  \href{http://arxiv.org/abs/0712.2822}{{\ttfamily arXiv:0712.2822 [hep-th]}}.

\bibitem{Gilmore:2009ea}
J.~B. Gilmore, A.~Ross, and M.~Smolkin, ``{Caged black hole thermodynamics:
  Charge, the extremal limit, and finite size effects},''
  \href{http://dx.doi.org/10.1088/1126-6708/2009/09/104}{{\em JHEP} {\bfseries
  09} (2009) 104}, \href{http://arxiv.org/abs/0908.3490}{{\ttfamily
  arXiv:0908.3490 [hep-th]}}.

\bibitem{Tolman:1930ona}
R.~Tolman and P.~Ehrenfest, ``{Temperature Equilibrium in a Static
  Gravitational Field},'' \href{http://dx.doi.org/10.1103/PhysRev.36.1791}{{\em
  Phys. Rev.} {\bfseries 36} no.~12, (1930) 1791--1798}.

\bibitem{Porfyriadis:2018jlw}
A.~P. Porfyriadis, ``{Near-$AdS_2$ perturbations and the connection with
  near-extreme Reissner{\textendash}Nordstrom},''
  \href{http://dx.doi.org/10.1140/epjc/s10052-019-7347-6}{{\em Eur. Phys. J. C}
  {\bfseries 79} no.~10, (2019) 841},
  \href{http://arxiv.org/abs/1806.07097}{{\ttfamily arXiv:1806.07097
  [hep-th]}}.

\bibitem{Anninos:2020hfj}
D.~Anninos, F.~Denef, Y.~T.~A. Law, and Z.~Sun, ``{Quantum de Sitter horizon
  entropy from quasicanonical bulk, edge, sphere and topological string
  partition functions},'' \href{http://dx.doi.org/10.1007/JHEP01(2022)088}{{\em
  JHEP} {\bfseries 01} (2022) 088},
  \href{http://arxiv.org/abs/2009.12464}{{\ttfamily arXiv:2009.12464
  [hep-th]}}.

\bibitem{Law:2025ktz}
Y.~T.~A. Law, ``{de Sitter horizon edge partition functions},''
  \href{http://dx.doi.org/10.1103/cm6p-9r4n}{{\em Phys. Rev. D} {\bfseries 113}
  no.~8, (2026) 086005}, \href{http://arxiv.org/abs/2501.17912}{{\ttfamily
  arXiv:2501.17912 [hep-th]}}.

\bibitem{Dias:2004px}
O.~J.~C. Dias and J.~P.~S. Lemos, ``{Pair creation of higher dimensional black
  holes on a de Sitter background},''
  \href{http://dx.doi.org/10.1103/PhysRevD.70.124023}{{\em Phys. Rev. D}
  {\bfseries 70} (2004) 124023},
  \href{http://arxiv.org/abs/hep-th/0410279}{{\ttfamily arXiv:hep-th/0410279}}.

\bibitem{Blacker:2025zca}
M.~J. Blacker, A.~Castro, W.~Sybesma, and C.~Toldo, ``{Quantum corrections to
  the path integral of near extremal de Sitter black holes},''
  \href{http://dx.doi.org/10.1007/JHEP08(2025)120}{{\em JHEP} {\bfseries 08}
  (2025) 120}, \href{http://arxiv.org/abs/2503.14623}{{\ttfamily
  arXiv:2503.14623 [hep-th]}}.

\bibitem{Shi:2026wnk}
X.~Shi, G.~J. Turiaci, and C.-H. Wu, ``{The fate of nucleated black holes in de
  Sitter quantum gravity},''
  \href{http://dx.doi.org/10.1007/JHEP08(2026)191}{{\em JHEP} {\bfseries 08}
  (2026) 191}, \href{http://arxiv.org/abs/2605.03015}{{\ttfamily
  arXiv:2605.03015 [hep-th]}}.

\end{thebibliography}\endgroup
\bibliographystyle{utphys}
\end{document}